\documentclass[aps,pra,epsfigure,twocolumn,longbibliography,superscriptaddress]{revtex4-1}

\usepackage{amsfonts}
\usepackage{amssymb}
\usepackage{amsmath}
\usepackage{calc}
\usepackage{subfigure}
\usepackage{graphicx}
\usepackage{epstopdf}
\usepackage{dcolumn}
\usepackage{bm}
\usepackage{color} 
\usepackage{txfonts}
\usepackage[dvipsnames]{xcolor}
\usepackage[colorlinks, citecolor=blue]{hyperref} 
\usepackage{ulem}

\newcommand{\beq}{\begin{equation}}
	\newcommand{\eeq}{\end{equation}}
\newcommand{\beqa}{\begin{eqnarray}}
	\newcommand{\eeqa}{\end{eqnarray}}

\newcommand{\mean}[1]{\left\langle #1\right\rangle}
\newcommand{\RNum}[1]{\uppercase\expandafter{\romannumeral #1\relax}}

\usepackage{amsmath}
\usepackage[ruled]{algorithm2e}  

\begin{document}

	\title{Machine-learning assisted  quantum control  in random environment}

	\author{Tangyou Huang}
	\affiliation{International Center of Quantum Artificial Intelligence for Science and Technology (QuArtist) 
	\\ and Department of Physics, Shanghai University, 200444 Shanghai, China}
	\affiliation{Department of Physical Chemistry, University of the Basque Country UPV/EHU, Bilbao, Spain}

   \author{Yue Ban}
	\affiliation{Department of Physical Chemistry, University of the Basque Country UPV/EHU, Bilbao, Spain}
	\affiliation{School of Materials Science and Engineering, Shanghai University, 200444, Shanghai, People’s Republic of China}

	\author{E. Ya. Sherman}
	\affiliation{Department of Physical Chemistry, University of the Basque Country UPV/EHU, Bilbao, Spain}
	\affiliation{IKERBASQUE Basque Foundation for Science, 48013 Bilbao, Spain}
	
	\author{Xi Chen}
	\affiliation{Department of Physical Chemistry, University of the Basque Country UPV/EHU, Bilbao, Spain}

	\date{\today }

	\begin{abstract}
		
Disorder in condensed matter and atomic physics is responsible for a great variety of fascinating quantum phenomena,
{which are still challenging for understanding, not to mention the relevant dynamical control.} 
Here we introduce {\it proof of the concept} and analyze neural network-based machine learning algorithm for 
achieving feasible high-fidelity quantum control of a particle in random environment. 
To explicitly demonstrate its capabilities, 
we show that convolutional neural networks are able to solve this problem as they 
can recognize the disorder and,  by supervised learning,  further produce the policy 
for the efficient low-energy cost control of a quantum particle in a time-dependent random potential.  
We have shown that the accuracy of the proposed algorithm is enhanced by a higher-dimensional mapping of the disorder 
pattern and using two neural networks, each properly trained for the given task. 
The designed method, being computationally more efficient than the gradient-descent optimization, 
can be applicable to identify and control various noisy quantum systems on a heuristic basis.
	
  \end{abstract}
	\maketitle
	
    \section{introduction}
	Machine learning (ML), which enables computers to learn automatically from available task-specific data  
	\cite{krizhevsky2012imagenet,lawrence1997face,mnih2015human,silver2016mastering}, 
	is {revolutionizing} modern approaches in physical sciences \cite{Goodfellow-et-al-2016}.
	In quantum science, ML 
	becomes useful and powerful  \cite{CarleoRMP2019} in particle physics,  many-body physics \cite{Carleo602}, 
	and quantum computing \cite{ThomasPRX18} among others. Recently developed 
	learning architectures \cite{SchmidhuberNN15} such as convolution neural networks (CNN), 
	having a considerable success in object detection and image classification, were  beneficial to classify
	phases of matter \cite{CarrasquillaNature2017}, study non-equilibrium glasses  \cite{ML2017pans}, 
	find hidden order in electronic-quantum-matter imaging data \cite{ML2019nature} and identify the thermodynamic 
	time arrow \cite{seif2021machine}.
	
	
   All the above studies were performed for systems where disorder is either nonexisting or 
   plays a negligible role in the system dynamics.  In practice, impurities, noise, and other imperfections are ubiquitous and 
   unavoidable in condensed matter \cite{Disorderedreview} 
   and its simulated counterparts \cite{sanchez2010disordered}. 
   Particularly,  the ultracold atoms offer a feasible and controllable platform for studying  
   the disorder 
   \cite{LyePRL2005,AspectPRL2005,FortPRL2005,roati2008anderson}.  
   In this scenario, 
   the random potential is implemented by optical means, and brings about a variety of 
   intriguing phenomena \cite{ShapiroPRL2007,DriesPRA2010,ChengPRA2010,VolchkovPRL18,YuePRA20}, 
   i.e. localization effects, phase transitions, and superfluidity, due to 
   the interplay among the disorder, nonlinearity, trapping potential or/and spin-orbit coupling. 
   Along with these 
   developments, the power of supervised learning (SL) 
   is harnessed to categorize stochastic data, extract quantitative information from this data,
   and predict the features of complex quantum systems, at a reasonable computational 
   cost \cite{ref_arxiv1802-04063,ref_arxiv1802-04063,pilati2019supervised,Guo2021,ohtsuki2020drawing,SaraceniPRE2020,PhysRevA.104.052412}. 

   However, quantum control under disorder 
   still remains a major challenge \cite{WuPRL2011,MardonovPRL15,ScoquartPRR20,ref_Neurocomputing,ref_npj5-33}, 
   though optimal control \cite{RabitzPRA98,CalarcoPRL11,ShersonPRA18}, 
   ML \cite{Henson13216,ref_arxiv1802-04063,BukovPRX2018,ref_npj5-85,ref_prl116},
   and shortcuts to adiabaticity  \cite{chenprl104,RevModPhys}
   have been exploited for fast manipulations
    in regular systems.  
   The extensive study of stochastic systems \cite{KosloffPRL,FunoPRL20}  
   have emerged in quest for controlling the dissipative 
   dynamics most efficiently. However, when it comes 
   to disorder, to classify or identify stochastic data embodied in the dynamics is a conundrum. 
   As the size of the stochastic sample increases dramatically, the higher power of ML is demanding in such complexity.
   
   To work out this problem,   we establish the ML approach for identifying and controlling dynamics of a quantum system 
   with disorder. For this purpose,
   we use deep learning with two CNNs for high-fidelity control of  a 
   quantum particle in a time-varying trapping potential embedded in random environment. 
   We begin with an important result: training the CNN can 
   efficiently preselect the relevant type of the disorder realization from 
   tens of thousands of stochastic samples. 
   Then, we introduce the second CNN to find the optimal control policy such as the time-dependent potential shape, 
   in a training regression model. To make the optimization more efficient, the randomness 
   classification from deep learning is an essential pretraining for disordered system under control,  
   thus removing the redundant data. 
   Thus, the SL with CNNs provides the ability to generalize the tasks beyond their original design,
   applicable to any realization of random potential. Our methods pave an efficient way 
   for the robust optimal control, i.e.  cooling, transporting, trapping atoms or 
   charged particles (ions and electrons) \cite{Henson13216,chenprl104,CiracPRX17}, by taking into account environmental 
   noise and randomness.
   
   \section{Disordered system and control strategy}
   Consider a quantum particle of mass $m\equiv\,1$, 
   located at the sum of time-dependent harmonic potential and a random potential of impurities. The corresponding
   Hamiltonian (with $\hbar\equiv\,1$) reads 
   \beq
   \label{H}
   H(t) = \frac{p^{2}}{2}+ \frac{1}{2}\omega^{2}(t)x^{2}+U_{r}(x),
   \eeq
   where $p$ is the momentum, $\omega(t)$ is the frequency of harmonic trap, and $U_{r}(x)$ is the random potential 
   of interest. Equation (\ref{H}) describes atoms in optical traps and
   electrons in  acoustic traps \cite{CiracPRX17} and gate-formed quantum dots. 
   The motivation behind the frequency modulation, i.e. from $\omega(t\leq 0) = \omega_{0}$ to $\omega(t=t_{f}) = \omega_{f}$, 
   is to achieve the fast high-fidelity expansion/compression 
   within a short time $t_{f}$, 
   beyond the adiabatic criteria \cite{chenprl104,Henson13216}. 

\begin{figure}
	\includegraphics[width=\columnwidth]{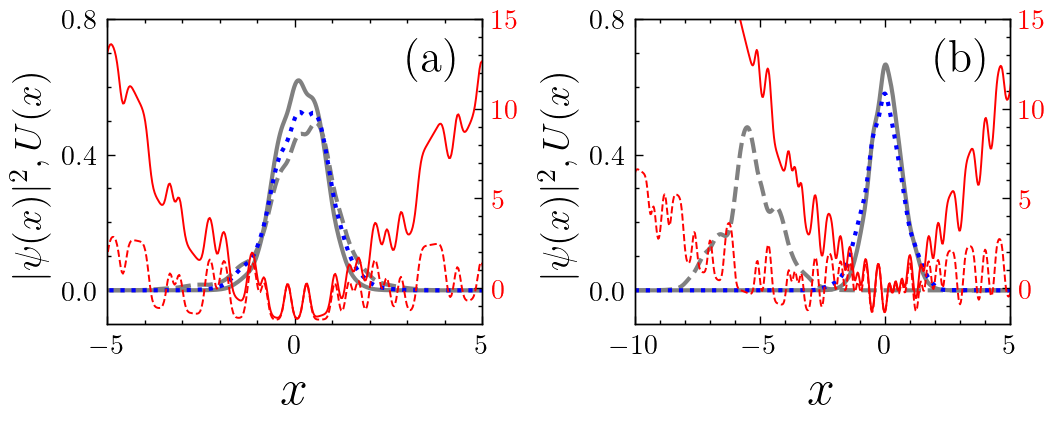}
	\caption{{Probability densities of the initial ($t=0,$ {black solid line}) and final ($t=t_{f},$ {black dashed line}) 
	 ground states in the harmonic trap in the random environment forming the total potential $U(x).$ 
	 Two realizations are presented to illustrate the effect of disorder. 
	 The corresponding final states (blue dotted lines) produced by the optimal control policy with SL
	 are shown as well. The total initial ({red solid line}) and final  ({red dashed line}) potentials, 
       are also shown for the eye. Parameters: $U_{0} = 1$, $\omega_{0} =1 $, and $\omega_{f} = 0.1$. Here and below we use  
    $\xi=d=1/8$ for ${\cal N}=160$ impurities at the $\{ -10, 10\}-$interval.}
 	{ 
 		Since we are using the system of units with $\hbar\equiv\,m\equiv\,1$, the length and the
 		energy are measured in the units of $1/\sqrt{\omega_{0}}$ and $\omega_{0},$ respectively.
 	}
	}\label{fig1}
\end{figure}

   We study generic random potential $U_{r}(x)$, corresponding to the Anderson-like disorder, produced by ${\cal N}\gg\,1$ impurities 
    at the positions $x_{j}=x_{j-1}+d$ regularly separated by the distance $d.$ The potential
   can be presented in the form: 
   \beq \label{Urx}
   U_{r}(x) = U_{0} \sum_{j=1}^{\cal N} s_{j} u(x-x_j).
   \eeq
   with $u(z) = \exp(-z^{2}/\xi^{2}).$ Here $U_{0}$ is the amplitude potential of a single impurity, and $s_{j} = \pm 1$ is a random function 
   of $j$ with mean values $\mean{s_{j}}=\mean{U_{r}(x)}= 0$, {and correlators 
   $\mean{s_{j}s_{l}}=\delta_{jl}$, 
   $\mean{U_{r}(x)U_{r}(x^{\prime})}=\sqrt{\pi/2}U_{0}^{2}\xi\exp\left(-(x-x^{\prime})^{2}/2\xi^{2}\right)/d.$ 
   Each disorder realization is a random sequence of $\pm 1$, e.g., $S_{i} [j]=\{1,-1,1...1\}$, with $i$ and $j$ being the realization number 
   and impurities position, respectively.}  

   We consider narrow impurities, where the width $\xi$ satisfies condition $\xi\ll U_{0}^{-1/2}$ and the corresponding localization 
   length at the impurity with $s_{j}=-1$ is of the order of $1/(U_{0}\xi) \gg d.$  
   Thus, localization by disorder involves many impurities \cite{MardonovPRL15} while the interaction energy with a single 
   impurity behaves as $\sim  U_{0}{\xi}s_{j}|\psi(x_j)|^{2},$ where $\psi(x)$ is the wavefunction. 
   For a sufficiently strong parabolic potential $\omega^{2}x^{2}/2$, the ground state
   has the energy close to $\omega/2$ and the harmonic oscillator width $w_{\rm ho}\sim 1/\sqrt{\omega}.$ 
   As the potential fluctuations behave as $\sqrt{N_{\rm imp}},$ where $N_{\rm imp}\sim 1/(d\sqrt{\omega})$ 
   is the number of impurities at the localization length of the state, 
   we estimate the shift in the ground state energy 
   as $\Delta\epsilon/\omega\sim U_{0}\xi/(\sqrt{d}\omega^{3/4}).$ To estimate the length $\ell$ of 
   the disorder-induced 
   localization, we minimize the sum of the kinetic energy $\sim 1/\ell^{2}$ and potential energy in 
   the disorder potential as $\sim U_{0}\xi/\sqrt{\ell d}$
   and obtain $\ell\sim (U_{0}\xi/\sqrt{d})^{-2/3}$ with the corresponding energy 
   $\epsilon_{\rm loc}\sim (U_{0}\xi/\sqrt{d})^{4/3}.$ 
   Therefore, in the parabolic potential, localized states can be located 
   at the distances up to  $w_{d}\sim\sqrt{\epsilon_{\rm loc}}/\omega\sim(U_{0}\xi/\sqrt{d})^{2/3}/\omega$, 
   meaning that with the decrease in $\omega,$ the ground state can be positioned at a 
   large distance from the origin.   
   
   Figure \ref{fig1} illustrates that the eigenstates of the final trap can be completely 
   changed by different realizations of random potential, as compared to the disorder-free results. 
   {For the realization in Fig. \ref{fig1} (b), where the initial and final states are almost orthogonal, the high-fidelity results cannot be achieved even with the optimal 
   control policy presented below.} 
   This intriguing feature makes the previous methods \cite{chenprl104,Henson13216}  
   {invalid} in our current problem.  As a consequence,  we need improved statistical analysis and computational method. 
    

	\begin{figure}
		\centering	
		\includegraphics[width=\columnwidth]{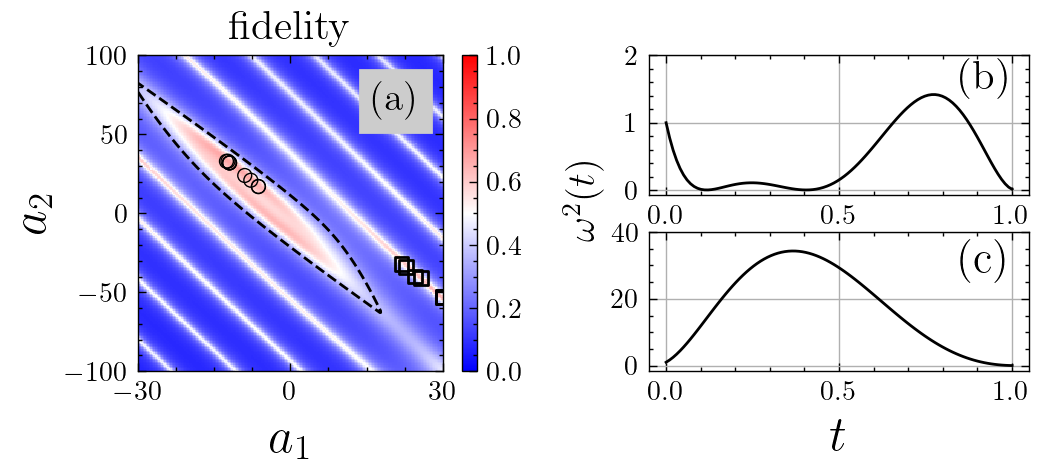}
		\caption{ (a) {The fidelity of the control policy $A=\{a_{1},a_{2}\}$ for 
		disorder-free harmonic potential (the blue-pink background) 
		with the high-fidelity zone (dashed line) satisfying the criteria 
		$\omega_{\max}^{2}(t)\leq \Omega^{2}.$ 
		The  ``feasible'' and ``unfeasible'' (with $F \geq F_{b} =0.9,$ where $F_{b}$ is the fidelity bound)
		control policies in the presence of disorder 
		are indicated by $``\circ"$ and $``\square"$ symbols. 
		Two example  functions of $\omega^{2}(t)$ are compared in (b) and (c), corresponding to the 
		''feasible`` and ''unfeasible`` solutions. Parameters: $\omega_{0} =1$, $\omega_{f}=0.1$, $t_{f}=1$, 
		and $\Omega= \sqrt{6},$ taken here as an example. Interestingly, for a given set $A=\{a_{1},a_{2}\}$ 
		the fidelity in the presence of disorder can be higher than that for the disorder-free harmonic potential.
		{The parameters $a_{1}$ and $a_{2}$ are measured in the units of $\omega_{0}^{2}$ and $\omega_{0}^{3},$
			respectively.} Note that behavior of $\omega^{2}(t)$ in (b) is counterintuitive since it includes 
			a considerable increase at $t$ close to $t_{f}=1.$
	}
	}
	\label{fig2}
	\end{figure}
	
	To {proof the principle of ML application} we choose the third-order polynomial 
	\beq
	\label{omega}
	\omega(t) = a_{0}+a_{1} t+a_{2}t^{2}+a_{3}t^{3},
	\eeq
	 as the control function for the trap frequency, where $a_{0}=\omega_{0}$, and 
	 $a_{3} = \left[\omega_{f}-(\omega_{0}+a_{1} t_{f}+a_{2}t^{2}_{f})\right]/t^{3}_{f}$ 
	 are given by the boundary conditions, 
	 $\omega(0)=\omega_{0}$ and $\omega(t_{f})=\omega_{f}$. The initial state at $t=0$ is
	 assumed to be the ground state for simplicity. 
	{The freedom left in $a_{1}$ and $a_{2}$ offers the possibility to optimize the control function $\omega(t)$, thus finding
	the maximum  ground-state fidelity defined as
\beq
          \label{Fidelity}
          F\equiv\left|\int_{-\infty}^{\infty}\psi^{*}(x,t_{f})\psi_{\rm gr}(x|\omega_{f})dx\right|^{2},
\eeq	
	 where $\psi_{\rm gr}(x|\omega_{f})$ is the ground state in the random potential 
	 corresponding to $\omega_{f},$  and  $\psi(x,t_{f})$ is obtained by a direct numerical solution 
	 of the non-stationary Schr\"{o}dinger equation with the Hamiltonian $H(t)$ from Eq. \eqref{H}. 
	 The optimal design of the trap frequency through the control policy $A=\{a_{1},a_{2}\}$ 
	 can produce $\psi(x,t_{f})$ with the maximum possible fidelity. 
	 
	 We impose two conditions on the optimal control function, 
	 {with the hint from the analysis on the high-fidelity control without disorder in Appendix \ref{AppendixA}}. First, it has to provide a high fidelity for the 
	 quantities of interest, in this case, as defined in Eq. \eqref{Fidelity}. Second, $\omega^{2}(t)$ 
	 should correspond to a moderate energy consumption required for the transition, 
	 suggesting that the maximum $\omega_{\max}^{2}(t)$ does not exceed a certain value $\Omega^{2}$ such that the 
	 process is experimentally feasible.
	 Figure \ref{fig2} illustrates the high-fidelity 
	 zone control policy $A=\{a_{1},a_{2}\}$ and corresponding {feasible} control function $\omega(t)$, satisfying the  
	 criteria $\omega_{\max}^{2}(t)\leq \Omega^{2} = 6$.
	 The search for the optimal coefficients in the {relevant  $\{a_{1},a_{2}\}$ range (see Fig. \ref{fig2}) 
	 is a time-consuming task even for a given realization.}
	 Since the stationary state and dynamics rely on the disorder realizations, the optimization of  
	 control policy 
	 also requires immense computing power.  { Note that the total number of disorder realizations in Eq. \eqref{Urx}
	 	is approximately  $2^{\cal N}.$ However, in agreement with the manifold hypothesis \cite{manifoldhyp}, 
	 	many of these realizations produce similar $U_{r}(x)$-functions 
	 	with similar $\psi_{\rm gr}(x|\omega_{f})$ width
	 	and positions. Therefore, the ML can use databases of moderate $(\le 10^{5})$ size.} 
 	 In what follows, we  {are motivated to} develop the SL
	 based on two CNNs to overcome such 
	 challenge.  

  \begin{figure}
	\centering
	\includegraphics[scale = 0.29]{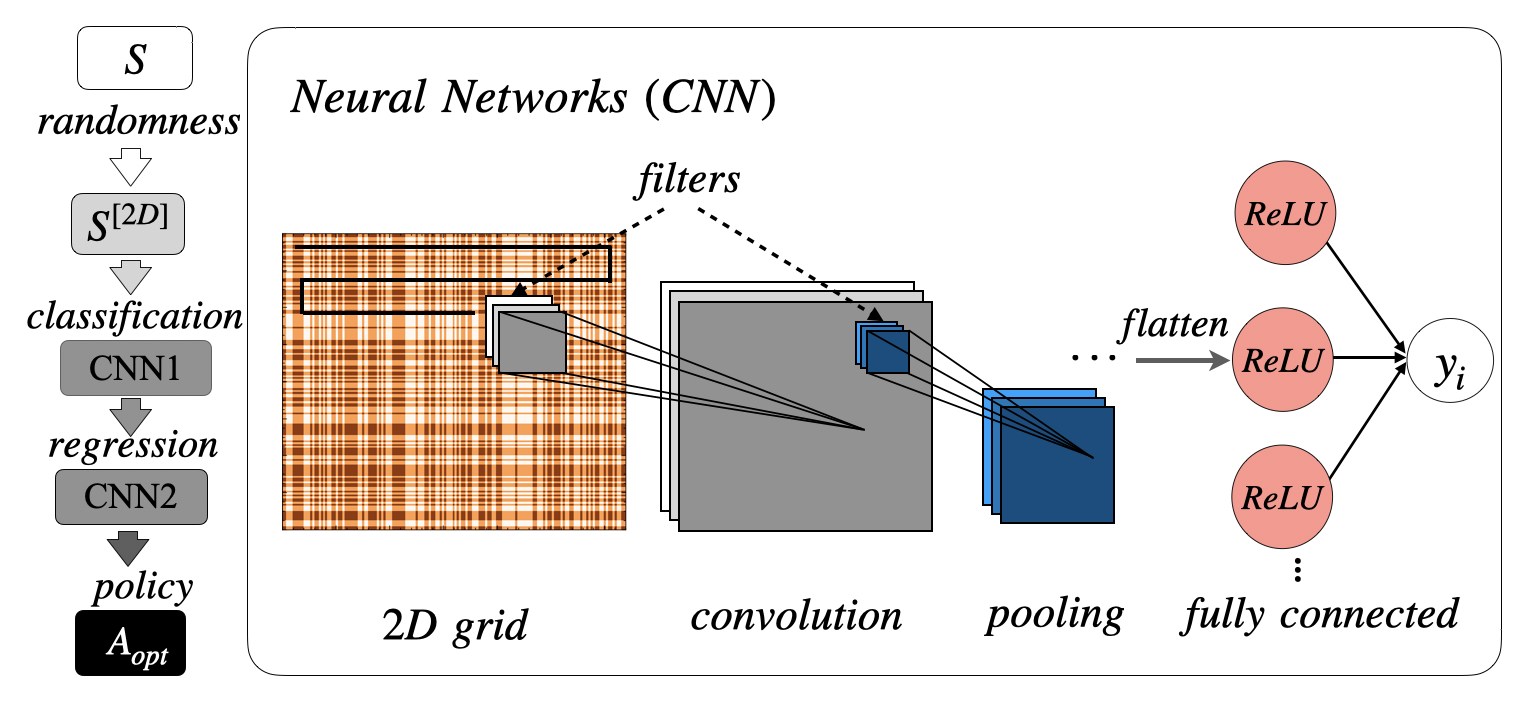}
	\caption{Schematic diagram (left) of SL with two CNNs for randomness recognition and regression.  
	Working flow (right) of CNN includes conversion from 1D $S_{i}[j]$ to 2D grid $S_{i}^{\left[2D\right]}\left[j_{1},j_{2}\right]$, 
		convolution and pooling layers, fully-connected layer with the ${\tt ReLU}$ activation function and the output $y_{i}$. 
		Details, including description of the ${\tt ReLU}$ function, are presented {in Appendix \ref{AppendixB}}. 
	}\label{fig3}
\end{figure}

  \section{Machine learning procedure} \label{Mlp}
   Now, we proceed to use SL, comprising two CNNs, for classifying the disorder realizations and constructing 
   the optimal control policy, 
   through the connection between the random sequence $S_{i}[j]$ and the optimal control policy $A_{\rm opt}$, 
   see the schematic diagram in 
   Fig. \ref{fig3}.  {One can refer to Appendix \ref{AppendixB1} for the technical description of SL.}
  
   First, we generate $4 \times 10^4$ disorder realizations with the labeled sequences $S_{i}[j]$ as the inputs. 
   For each realization, the fidelity of overlap between the eigenstates at $t=t_{f}$ and final wavefunctions resulting 
   from the state evolution [see Eq. \eqref{Fidelity}] is numerically 
   calculated with the control function $\omega(t)$ in Eq. (\ref{omega}). The maximum 
   fidelity for the given $i-$the realization, $F^{\max}_{i},$ and corresponding control policy $A_{i}$ are thus determined by using the same approach in Fig. \ref{fig2}, 
   where the criteria $\Omega^{2}=6$ and $F_{b} =0.9$ are applied to bound the {feasibility and fidelity},  
   while keeping a considerable size of database. The whole database $X=\{S_{i}, F^{\max}_{i}, A_{i}\}$ 
  is finally established, 
   where $80 \%$ of the database is selected as a training set, and the rest as a testing set.

   \begin{figure*}
   	\centering
   	\includegraphics[scale = 0.85]{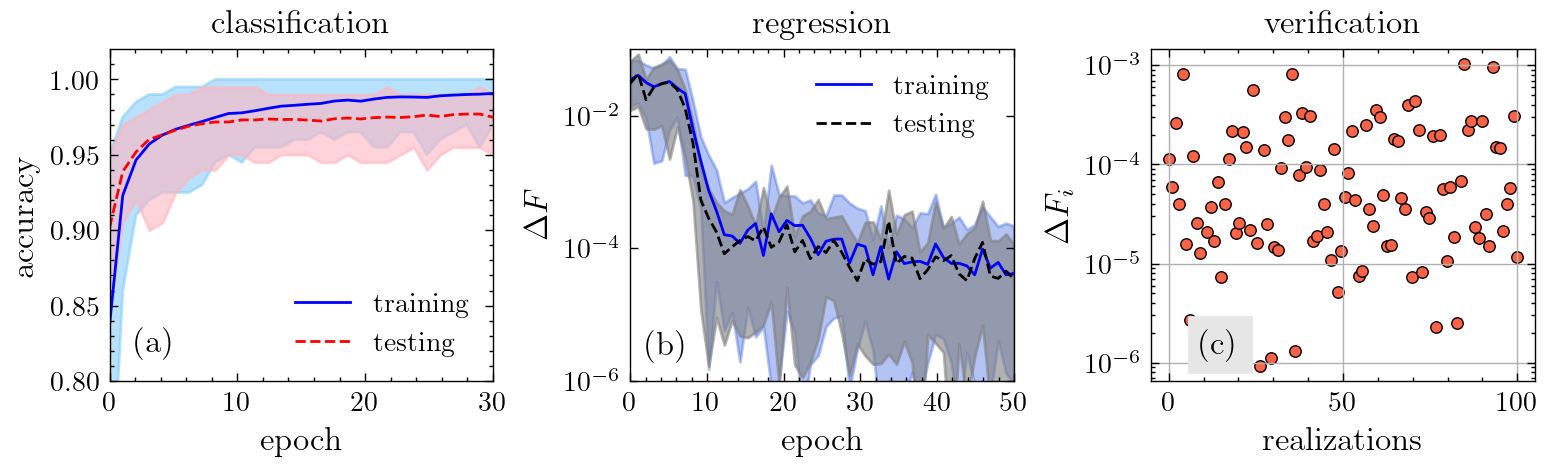}
   	\caption{ 
   		 {The accuracy of CNN1 (a) and the fidelity deviation (b) 
   		 	   	are displayed for classification and regression, where the dashed and 
   		 	   	solid lines represent the average value of test and training batches in each epoch.
   		 	The shadow area indicates the value distribution of batches. 
		(c) The fidelity deviation from two trained CNNs  are presented for $100$ testing realizations of random potential.}
   	}
   	\label{fig4}
   \end{figure*}

 Then, we introduce the first CNN, named in what follows CNN1, 
 in deep learning to assign each given realization of the random potential to a set of classes, 
 for instance, whether it determines {feasible} high-fidelity (FH) or not. Such randomness recognition is  classification, aiming at 
 selecting the reasonable inputs of realizations.
  To be more efficient, we extend the input $S_{i}[j]$ into two-dimensional (2D) grid (see Appendix \ref{AppendixB2}) 
  before the neural network is trained, by converting each sequence $S_{i}[j]$ 
  into a two-dimensional (2D) matrix $S_{i}^{\left[2D\right]}\left[j_{1},j_{2}\right]$ by using 
  \beq \label{2d_sequence}
S_{i}^{\left[2D\right]}\left[j_{1},j_{2}\right] \equiv S_{i}\left[j_{1}\right]+S_{i}\left[j_{2}\right].
\eeq 
 As expected,  the randomness recognition based on 2D grid surpasses the one-dimensional (1D) one, 
 in the sense that  the accuracy of classification and loss of regression are improved at the cost of computation time. 
 Therefore, we use $S_{i}^{\left[2D\right]}\left[j_{1},j_{2}\right]$ 
 as the inputs and $y_{i}$ as the output, where FH ($y_{i}=1$) and anti-FH ($y_{i}=0$) 
 suggests the aforementioned criteria, $F> F_{b}=0.9$ and $\omega_{\max}^{2}(t)\leq\,6$, is satisfied or not.  

 For classification in the CNN1 we employ the standard sequential structure (convolution and pooling layers),  
 and choose the loss function as  ${L_{1}(y,p)= -\sum_{i} \left[ y_{i} \log(p_{i})\right]} $, 
 with  ${p_{i}}$ being the probability produced by network, and the accuracy $N_{r}/N$, with $N_{r}$ being the 
 number of the right predictions out of total $ {N}$. 
 After  using optimizer ${\tt Adam()}$ at the rate of $10^{-4}$,  we manage 
 to select $5886$ out of $4 \times 10^{4}$ realizations, 
 with the accuracy above $97 \%$, {see Fig. \ref{fig4} (a) and the relative portion of the selected realizations
 being of the order of the $w_{\rm ho}/w_{d}\sim\omega_{0}^{1/2}/(U_{0}\xi/\sqrt{d})^{2/3}$ ratio, where $w_{\rm ho}$ is taken at $t=0.$}  Obviously, 
 this pretraining process is critical for classifying the disorder and excluding realizations yielding the 
 low-fidelity control, as shown in Fig. \ref{fig1} (b). 
 {Remarkably, the high efficiency of CNN1 can be conceptually interpreted by comparing 
 its feature-map with the corresponding position of final wave packet,  
{also see the detailed discussion in Appendix \ref{AppendixC2}.} 
  
Next, to find the optimal control policy $A_{\rm opt}$, we construct the second CNN (CNN2) for regression.
We choose the loss function  $L_{2}(y,y^{\prime}) = \sum_{i}(y_{i}-y_{i}^{\prime})^{2}/N$, where  $y$ and $y^{\prime}$ 
are the actual and predicted results of control policy $A$. The residual neural 
network \cite{ResNet} is used in CNN2, with a shortcut channel.  
We define fidelity deviation of each realizations $\Delta F_i= |F^{\max}_{i} - F^{\prime}_{i}|$ with $F^{\prime}_{i}$ being the fidelity predicted 
by control policy. 
During the training process, we further define the average value over each $N$-sized batch $\Delta F = \sum_i^{N} \Delta F_i /N$ in every training epoch for quantifying the performance of CNN2. 
As a consequence,  we train the CNN2 for 
achieving $\Delta F  \leq  10^{-4}$, {see Fig. \ref{fig4} (b).} 
Thereby, during the process we record the loss at each batch and 
the fidelity of predicted policies, and finally obtain the trained CNN1 
and CNN2, 
as indicated by solid lines in Fig. \ref{fig4} (a, b). 
{Moreover, we produce 100 realizations for verifying the
	performance of trained CNNs in Fig. \ref{fig4} (c), and also discuss the dependence of 
	their efficiency on the hyperparameter 
	in {Appendix \ref{AppendixB3}}.
	 Accordingly, after training two CNNs with $4 \times 10^{4}$ input disorder realizations, 
	the optimal control policy $A_{\rm opt}$ to
	design $\omega(t)$  for the high-fidelity control 
	with any random potential is obtained.}


%
%

	\begin{figure}
	\centering
	\includegraphics[width=\columnwidth]{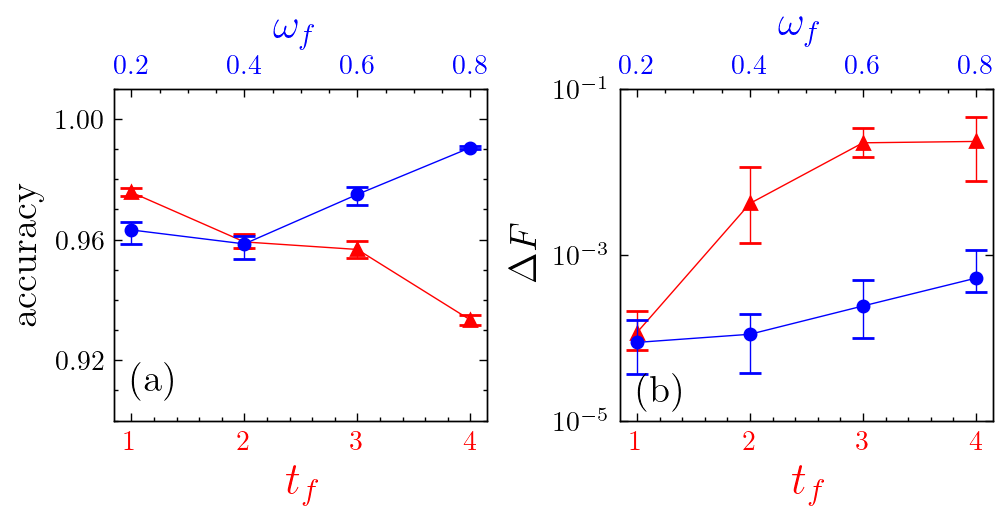}
	\caption{ The average accuracy in CNN1 (a) and the average fidelity deviation in CNN2 (b) 
	for the last ten epochs are illustrated for different $\omega_{f}$ and $t_{f}$,
	where the structure and hyperparameters are the same as those in Fig. \ref{fig4}, 
	and the error bars represent their deviations.
	}\label{fig5}
\end{figure}

\section{Discussion}

There are several points to be addressed on the generality of our proposed method. We can, in principle, choose  other $\omega(t)$ ansatzes 
with more parameters or even use the results from the gradient-descent optimization.
The detailed analysis clarifies that the influence of the form of ansatz 
(or moderate changing in the bound $F_{b}$ and/or $\Omega$) 
on the classification of disorder, performed by the CNN1, is essentially negligible, since the 
border line between high and low fidelity is mostly determined by the intrinsic property such as 
the shape of the disorder rather than by the external condition.
However, malfunctioning or poor performance of CNN1 can cause low efficiency of CNN2, obtaining the input from CNN1. 
The ansatz (\ref{omega}) serves as a reference  for setting the criteria.  
Note that  the CNNs trained with the gradient-descent optimization is not better than the ones with such 
simple ansatz, {see the detailed discussion in the Appendix \ref{AppendixC1}}.
	
Moreover, we can also apply the trained CNNs to different values of $t_{f}$ and $\omega_{f}$. Figure \ref{fig5} indicates 
the  average accuracy in CNN1 and the average fidelity deviation in CNN2 for the last ten epochs  
by using the same structure and hyperparameter as before. 
{On the one hand, when $\omega_{f}$ is increased, the 
random realizations are much easier to recognize, thus resulting in higher accuracy. 
It makes sense that the influence of random potentials on the fidelity can be negligible, 
when the  trap potential is strong enough to localize the state near the origin.}
However, the more realizations as
the inputs of CNN2 finally lead to the larger fidelity deviation as shown in Fig. \ref{fig5}. 
{On the other hand, according to the time-energy trade-off, 
larger $t_{f}$ (still far away from the adiabaticity) increase the area corresponding to condition 
$\omega_{\max}^{2}(t)\leq\Omega^{2}$ (cf. Fig. \ref{fig2}).
Thus,  more random realizations corresponding to the feasible $A_{\rm opt}$ increase the statistical
uncertainty  and degrade the performance of trained CNNs.} That is,  the fidelity deviation in CNN2  becomes larger 
because of worse classification,  depending on the distribution and number of the selected realizations in CNN1, see Fig. \ref{fig5}.  
In a word,  the combined effects of the trapping potential and disorder 
plays an important role in dynamical control, characterized by the fidelity and the 
required energy cost, e.g. the laser power for optical trap or the electrical power for quantum dots. 

 \section{Conclusions}
The behavior of quantum objects such 
as atoms and charged particles 
in random potentials is 
an active research area, with a lot of the accumulated knowledge and 
even more yet unknowns. 
The complexity  prevents the researchers from efficiently controlling the 
quantum dynamics in random environments. We presented a remedy  by developing {\it proof-of-principle}
supervised learning algorithms, trained through deep neural networks, 
to classify the randomness and find the optimal control policy. The efficiency and accuracy of the proposed algorithm 
is based on using two-dimensional mapping of the random potential and sequential
application of two neural networks, each trained for the given different task. 
Our results indicate that machine learning, based on the convolutional neural network 
for classification and regression, can be used to 
control various quantum systems with impurities, noise and imperfections, 
and ultimately to unveil the physical insight into the interplay of disorder 
and quantum dynamics. {With the advent of techniques of configurable optical 
traps \cite{optica2016} and surface acoustic waves \cite{hermelin2011electrons,mcneil2011demand}, 
we suggest the experimental verification of the proposed method for trapped atoms or electrons in random environment.} 

\section*{Acknowledgment}
This work is supported from 
NSFC (12075145), STCSM (2019SHZDZX01-ZX042019 and 20DZ2290900), SMAMR (2021-40), 
Program for Eastern Scholar, CSC fellowship (202006890071)
Basque Government IT986-16, PGC2018-095113-B-I00,  PGC2018-101355-B-I00 funded by MCIN/AEI/ 10.13039/501100011033 and by “ERDF A way of making Europe”,
EU FET Open Grant Quromorphic (828826) and EPIQUS (899368), and the Ramon y Cajal program (RYC-2017-22482). 



\appendix

\section{high fidelity quantum control without disorder}
\label{AppendixA}

To begin, we consider the case without random potential, $U_{r}(x)\equiv\,0$, in order to have a 
reference for understanding the effects induced by the disorder.
By setting $m \equiv 1$ and $\hbar \equiv 1$,
the  Hamiltonian of a single particle trapped in a harmonic potential reads
\beq
H = \frac{p^{2}}{2}+ \frac{1}{2} \omega^{2}(t)x^{2},
\eeq
which describes the compression and decompression 
by tailoring the  frequency $\omega(t)$ of the harmonic trap.
According to the Lewis-Riesenfeld invariant theory, 
the solution of time-dependent Schr\"{o}dinger equation admits  analytical expression \cite{chenprl104}:
\beq
\label{psi_ho}
\psi(x,t) = \left(\frac{\omega_{0}}{\pi b^{2}}\right)^{1/4}\exp\left[-\frac{i}{2} \int_{0}^t \frac{\omega_{0}}{b^{2}}dt^{\prime}\right]
\exp\left[i\frac{1}{2}\left(\frac{\dot{b}}{b}+i\frac{\omega_{0}}{b^{2}}\right)x^{2}\right],
\eeq
where the auxiliary function $b(t)$ satisfies the Ermakov equation:
\beq\label{ermakov}
\ddot{b}+\omega^{2}(t)b=\frac{\omega_{0}^{2}}{b^{3}}.
\eeq
For a decompression process from the initial frequency $\omega_{0}$ to final frequency $\omega_{f}$,  
the boundary conditions can be formulated as  $b(0)=1$, $b(t_{f})=\gamma$ ($\gamma=\sqrt{\omega_{0}/\omega_{f}} >1$), $\dot{b}(0)=\dot{b}(t_{f})=0$.
%
Thus, for an arbitrary control function $\omega(t)$,  we are able to calculate the time-dependent scaling parameter of $b(t)$ and 
corresponding $\dot{b}(t)$ by solving the Ermakov equation.
By considering the ground state, with the initial and final boundary conditions, we, in general, can reach the ideal target 
state, that is, $\psi_{\rm gr}(x|\omega_{f})= (\omega_{0}/\pi \gamma^{2})^{1/4} e^{-\omega _{0}x^{2}/(2\gamma^{2})}$.
{Based on  Eq.~(\ref{Fidelity})} , the fidelity can be analytically expressed as
\beqa
\label{F_expre}
F
=\left[\frac{4\omega_{0}^{2}b_{f}^{2}\gamma^{2}}{\omega_{0}^{2}(\gamma^{2}+b_{f}^{2})^{2}+(\dot{b}_{f}b_{f}\gamma^{2})^{2}}\right]^{1/2},
\eeqa
where $b_{f} = b (t_{f})$ and $\dot{b}_{f}=\dot{b}(t_{f})$ 
are the numerical solution of Eq. (\ref{ermakov}) at $t=t_{f}$.
Obviously, the fidelity $F$  strongly depends on  $b_{f}$ and $\dot{b}_{f}$.
When $b_{f} = \gamma$ and $\dot{b}_{f} =0$, we will have $F=1$. In this case, 
we recall the concept of shortcuts to adiabaticity, that is, to achieve fast 
adiabatic-like decompression without final excitation. 

\begin{figure}
	\centering
	\includegraphics[width=\columnwidth]{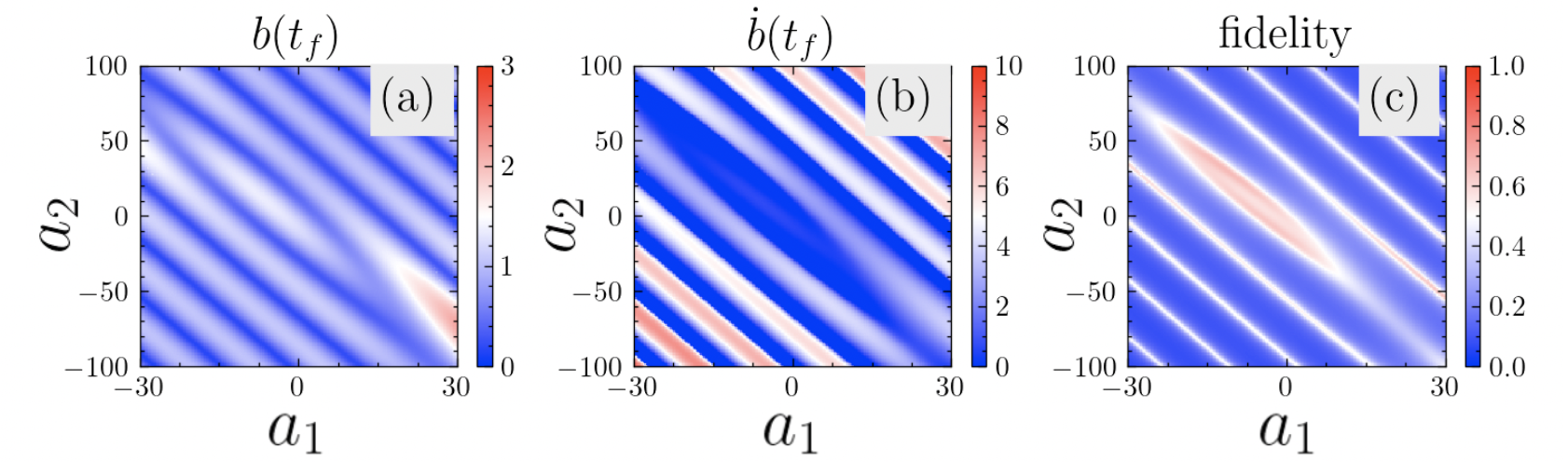}
	\caption{Dependence of $b(t_{f})$, $\dot{b}(t_{f})$ and the fidelity $F$ on
		the coefficient grid $\{a_{1},a_{2}\}$, where the parameters are $\omega_{0}=0$, $\omega_{f}=0.1$, and $t_{f}=1$.
		The straight lines corresponding to the maximum fidelity can be obtained with an approximate solution
		of Ermakov equation \eqref{ermakov} as $a_{2}=-3a_{1}/t_{f}+12n\pi/t_{f}^{3},$ with integer $n.$}
	\label{figap1}
\end{figure}

Without loss of the generality,
we choose the simple ansatz $\omega(t)= a_{0}+a_{1}t+a_{2}t^{2}+a_3t^{3}$, such 
that the fidelity is calculated, depending on the coefficients $a_{1}$ and $a_{2}$.
To understand the performance of the fidelity, 
Fig. \ref{figap1} (a) and (b) illustrate the 
dependence of $b(t_{f})$ and $\dot{b}(t_{f})$ on $\{a_{1},a_{2}\}$, respectively.
Figure \ref{figap1} (c) finally shows 
the plot of the fidelity dependent of the coefficient set $A =\{a_{1},a_{2}\}$,
in which the stripes occurs due to the interplay between $b(t_{f})$ and $\dot{b}(t_{f})$.
The high-fidelity regime in Fig. \ref{figap1} gives the criteria for machine learning later, when the random potential is involved.

\section{machine learning}
\label{AppendixB}

Machine Learning (ML) methods, including support vector machines, decision trees, random forests and artificial 
neural networks (ANNs), have been developed in last decades.
Moreover, the deep learning is proposed to handle  the huge quantity of data and complex system, notably,  
the ANNs is usually outperform than others.
Nowadays, the  ANNs are dedicated to solving complex tasks such as the image and video recognition, 
analysis of strategical games (AlphaGo), etc. 
In particular, the deep convolutional neural networks (CNNs), initially proposed for 
computer vision learning, now are overwhelming in the artificial intelligence (AI) industry.
Their unique architecture, inspired by research on the brain's visual cortex, greatly enhances
the performance of analysis of systems in complex surroundings, which is consistent 
with our problem on quantum control in a random environment.

{The reasons for using the CNNs to analyze the disordered system are three-fold}:
(1) Data grows exponentially with tremendous amount of disorder realizations;
(2) Training an ANN can be accelerated by using graphic processor units (GPUs);
(3) CNN can be used to identify the disorder as the application in image classification.

Next, we shall exploit the supervised learning, based on two CNNs, for classifying and controlling 
the joint effect of a regular (parabolic) potential and disorder. 	


\subsection{neural network  and supervised learning }
\label{AppendixB1}

A deep ANN consists of input, hidden and output layers, and the depth of network usually depends on the amount of hidden layer.
Meanwhile, a single layer is composed by a set of nodes, and each node is connected with the others  from the next layer with a particular weight and bias.
Moreover, the learning process of ANNs is combined by the forward-propagation and back-propagation computation based on the gradient descent algorithm.
We start with the propagating data from the input layer, pass the hidden layer(s), measure the output layer, and finally calculate network error based upon the network predictions.
With the error function and the gradient-base optimizer,  the back-propagation decreases error by updating the weights and bias of network.
Compared  with a regular ANNs,  the CNNs are trained to optimize the filters (or kernels) through the automated learning, instead of the hand-engineered in feature extraction.
It takes advantage of the hierarchical pattern in capturing data feature and reducing the number of the parameters involved.
In order to explain the functioning of this CNN, we shall make use of the following notation:

1. $ x^\ell$ is the data flow of $\ell^{th}$ layer.


2. The filter $K$ with the size $k_{1}\times k_{2}$ has $m$ and $n$ as the iterators.

3. The weight between $\ell$ layer and $\ell-1$ layer is represented by $\omega^\ell$, and the corresponding bias $b^\ell$.

4. $f(\cdot)$ is an activation function.

5. The underlying data of layer is $x_{i,j}^{\ell} = \sum_{m,n} f(w_{m,n}^{\ell}x_{m,n}^{\ell-1}+b^{\ell})$, where  $i$ and $j$ are the iterator.

6. $ x^{\ell}\otimes K^k$ represents the data extracting process by the $k^{th}$ filters.

7. $y_{i}$ and $y_{i}'$ are the actual and predicted values (labels), respectively.

Supposing that we use $k$ filters,  the output of $\ell$-th convolutional layer can be presented as
\beq
x^{\ell}_{i,j} = \sum_{k=0}^k x_{i,j}^{\ell-1}\otimes K^k= \sum_{k=0}^k\sum_{m,n}^{k_{1},k_{2}} f(K_{m,n}^{k}x_{i+m,j+n}^{\ell-1}+b^{\ell}),
\eeq
where the activation function $f(\cdot)$ is  the logistic ${\tt Sigmoid}$ function, $ f(z) = 1/(1+\exp(-z)) $,  or the rectified linear unit 
(${\tt ReLU}$) function, $f(z) = \max(0,z)$.
The ${\tt Sigmoid}$ function maps the data from $[-\infty,+\infty]$ into $[0,1]$, resulting in the probability of prediction
as the output of network.
And  the ${\tt ReLU}$ is a piecewise step function, ${\tt Relu(x)=\max(x,0)}$, that transfers the input data from $[-\infty,+\infty]$ into $[0,+\infty]$.
Such two nonlinear activation function{s} are widely used to  allow the nodes to learn more complex structures in the data. 
A pooling layer, aiming to reduce the spatial size, contains {\tt MaxPooling()} and {\tt AveragePooling()}.
More specifically, they extract the maximum (or average) value of the pooling block from the previous layer, thus reducing the amount of the parameters. 
The CNN layer is schematically shown in Fig. \ref{figap2}, in which we set the $ 16\times 16$ inputting data and three $7\times7$ filters for the convolution layer and three $2\times2$ filters for calculating the maximal pooling.

\begin{figure}
	\centering
	\includegraphics[width=\columnwidth]{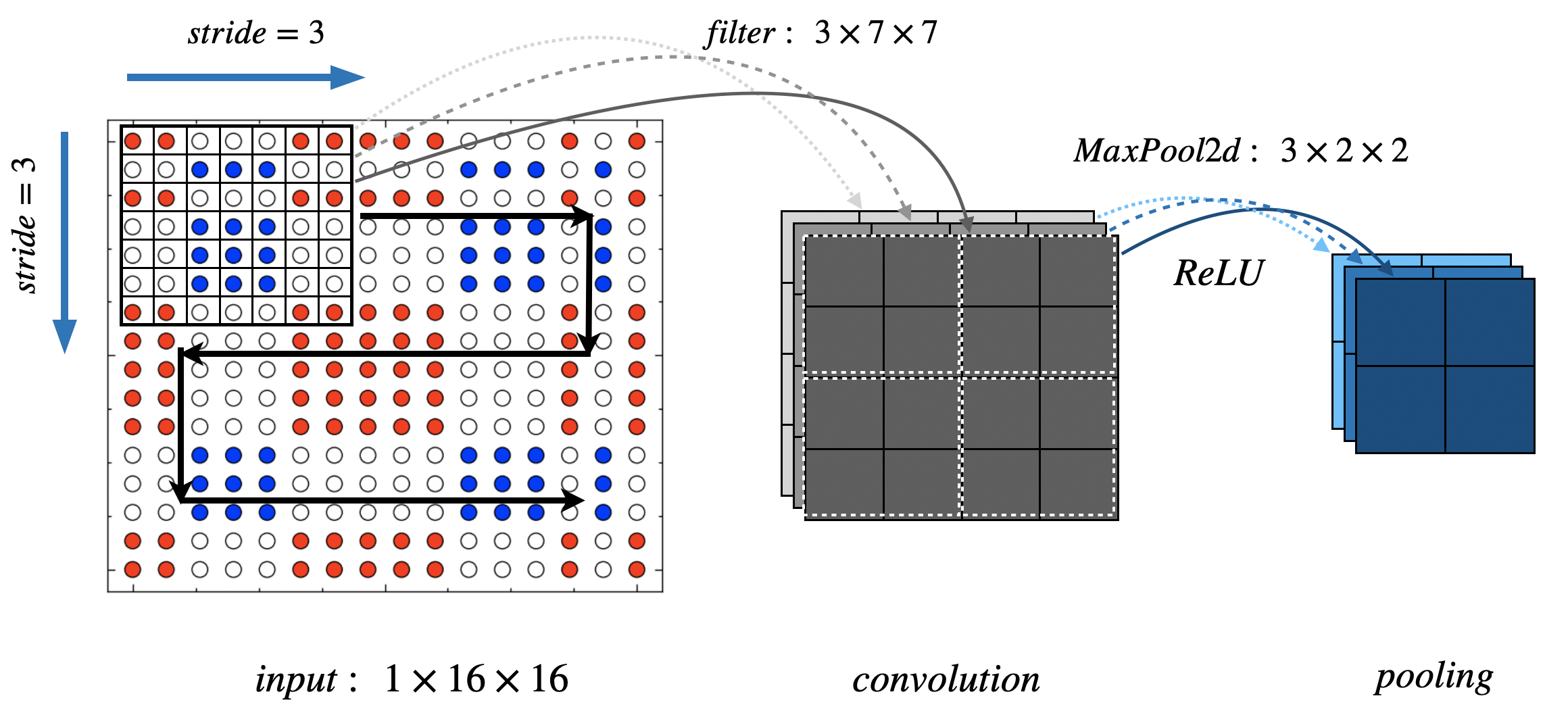}
	\caption{A single unit of CNN includes the convolution, activation and pooling process. 
	We take $16\times16$ grids as an example for illustrating the working flow and variables in the function {\tt Conv2d()} and {\tt Maxpool()}. 
	In this case, the process can be represented by {\tt Conv2d(1,3,7,3)} and {\tt MaxPool2d(2)} in the PyTorch. }
	\label{figap2}
\end{figure}

Next, we introduce the loss function and gradient-based optimizer for classification and regression.
Regarding the classification task,  the loss function is defined as the following cross-entropy form: 
\beq
\label{cost_class}
\mathcal{J}(W,b;y,y^{\prime})=  \frac{1}{N}\sum^{N}_{i =1}J_{1}(W,b; y_{i},y^{\prime}_{i}),
\eeq
with 
\beq
J_{1}(W,b; y_{i},y^{\prime}_{i})  = - y_{i} \log\left[\sigma (y^{\prime}_{i})\right],
\eeq
where $W$ is the weight collection of network for $N$ samples, and $\sigma (y^{\prime}_{i})$ is the softmax probability, where the 
Softmax function $\sigma (z_{i})= e^{z_{i}}/(\sum_{j}e^{z_j})$ is used for normalizing the output. 
As for the two-category image classification $ k = 2$ task, the input layer is a 
flatten pixel sequence $x_{i}$ of image, and the result is the probability of labels.
For instance, when the actual binary label is $y_{0} =\{1,0\}$, and two dimension output $y^{\prime}_{0} =\{p_{0},p_{1}\}$, 
the error for  a single prediction thus is
$j= - y_{0} \log[y_{0}']^{T} $.
On the other hand, for the regression process, the loss function in (\ref{cost_class}) is a mean squared error:
\beq\label{cost_regress}
J_{2}(W,b; y_{i},y^{\prime}_{i})= \sum_{i}|y_{i}-y^{\prime}_{i}|^{2},
\eeq
which represents the deviation from the regression prediction $y^{\prime}_{i}$ to the actual sample $y_{i}$.
We use the optimizer {\tt Adam()}, which is included in the 
application programming interface (API) of PyTorch, for optimizing the loss function in the 
learning process.
Back-propagation (or forward pass) refers to the calculation and storage of the intermediate variables 
(weights and bias) of a neural network, and minimizes the cost function by gradient-based optimizer. This can be simply expressed as 
\beq
{\tt Repeat}: \left\{W^{\ell}_{i,j}= W^{\ell-1}_{i,j}-\eta \frac{\partial \mathcal{J}}{\partial W^\ell_{m,n}}\right\} , 
\eeq
with learning rate $\eta$.

\begin{figure}
	\centering
	\includegraphics[width=\columnwidth]{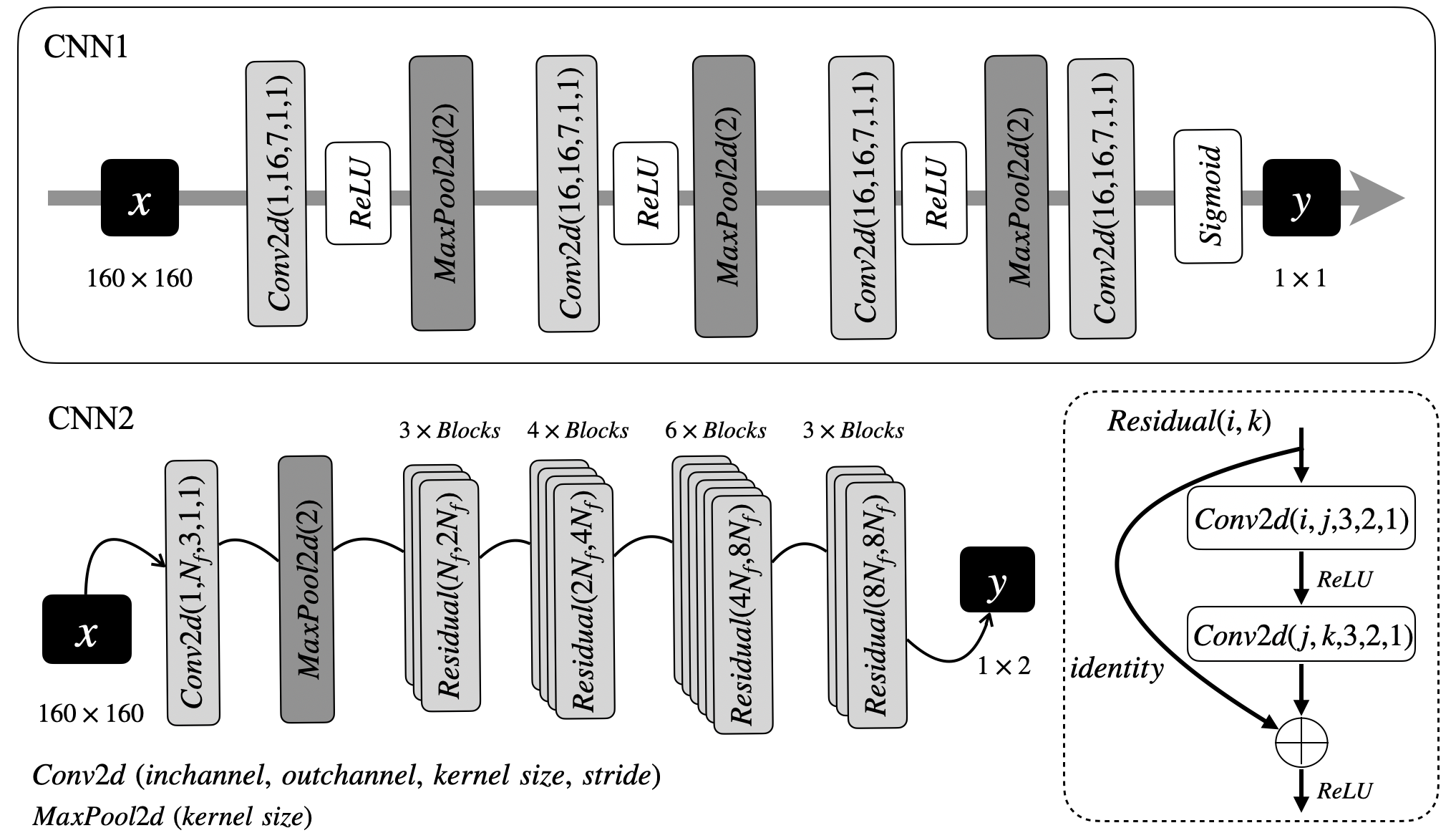}
	\caption{Diagrammatic architectures of CNN1 and CNN2, are illustrated, where the function and its parameters are presented for each layer of network and the residual block of CNN2 in the dashed frame is specified.
		More details can be found in the main text.
	}
	\label{figap3}
\end{figure}

Following that, we create the algorithm for our task, which consists of two CNNs for classification and regression, respectively.
We encode the algorithm based on the PyTorch \cite{pytorchtutorial} software platform, where the deep learning library 
consists of the tensor flow and the computation is accelerated by GPUs. 
In order to illustrate the learning algorithm, we briefly introduce the functions that we used 
in the PyTorch API:

1. 2D convolution layer: \\ ${\tt Conv2d~ (inchannel, outchannel, kernel size, stride)}$.

2. Max pooling layer ${\tt MaxPool2d~(kernel size)}$.

3. ${\tt ReLU}$ and ${\tt Sigmoid}$ represent the rectified linear unit function and the corresponding logistic function, respectively.

4. ${\tt CrossEntropyLoss()}$ and  ${\tt MSELoss()}$ indicates the loss function of Eqs. (\ref{cost_class}) and (\ref{cost_regress}).

The variables include:
$inchannel$: the depth of channel in the input,
$outchannel$: the number of output channel depends on the amount of filter (or kernel),
$kernel size$: the filter size,
$stride$:  controlling the stride for the cross-correlation.

The detailed parameters can be further found in the PyTorch tutorial \cite{pytorchtutorial}.
Along with this user-friendly platform, we now construct the algorithm for the supervised learning.  
Before proceeding, we should design the architecture of  CNN1 and CNN2, since the performance of a neural network mostly depends on its structure and layer depth.
According to the complexity of task, the architecture of CNN1 is built up as a standard sequential network and CNN2 as a $ResNet$ network \cite{S1ResNet},
see the details in the flow chart in Fig. \ref{figap3}.
The residual block ${\tt Residual()}$, with so-called ``identity shortcut connection", skips two layers, as shown in Fig. \ref{figap3}.
It makes the network possible to train hundreds layers,  keeping  the compelling performance.
After introducing the CNNs-based supervised learning and the architecture of two networks,  we can start with creating the  database and  training  the model for classification and regression.

\subsection{classification and regression}
\label{AppendixB2}

For supervised learning, two essential steps, including
data preparation and model training, are required. 
In this sense, the performance of model can be improved 
by increasing the training data and selecting a high-quality database. 
To calculate the database, however, is a time-consuming task for a complex system, so it 
is significant to preselect for producing a representative database with high quality. 
Let us consult the Fig. \ref{fig1} of the main text, in which 
the eigenstates of the final trap can be completely changed by different realizations of random potential, 
and some of them will results in the low-fidelity control for sure. 
Thus, we propose CNN1 for the preselection, in order to establish the link between input and output data of 
network  by choosing a small amount of high-quality database. We will demonstrate that the high-quality database not only 
brings the benefits to training process, but also makes the
trained network more universal and tolerant.

Aiming to present the feature of each single random sequence, we initially extend the one-dimension (1D) sequence in two-dimension (2D) grid, see Eq.~(\ref{2d_sequence}). 
More specifically, as in the main text, we select a $1\times160$ random sequence, e.g., $S_{i}[j] = \{1,1,-1,1,...,-1,1\}.$  
A typical resulting 2D grid with the elements 2,0, and $-2$, is shown in Fig. \ref{figap4} (d).
We will see the advantage of 2D $S_{i}^{\left[2D\right]}\left[j_{1},j_{2}\right]$ as the input data, in the following discussion.

\begin{figure}
	\centering
	\includegraphics[width=\columnwidth]{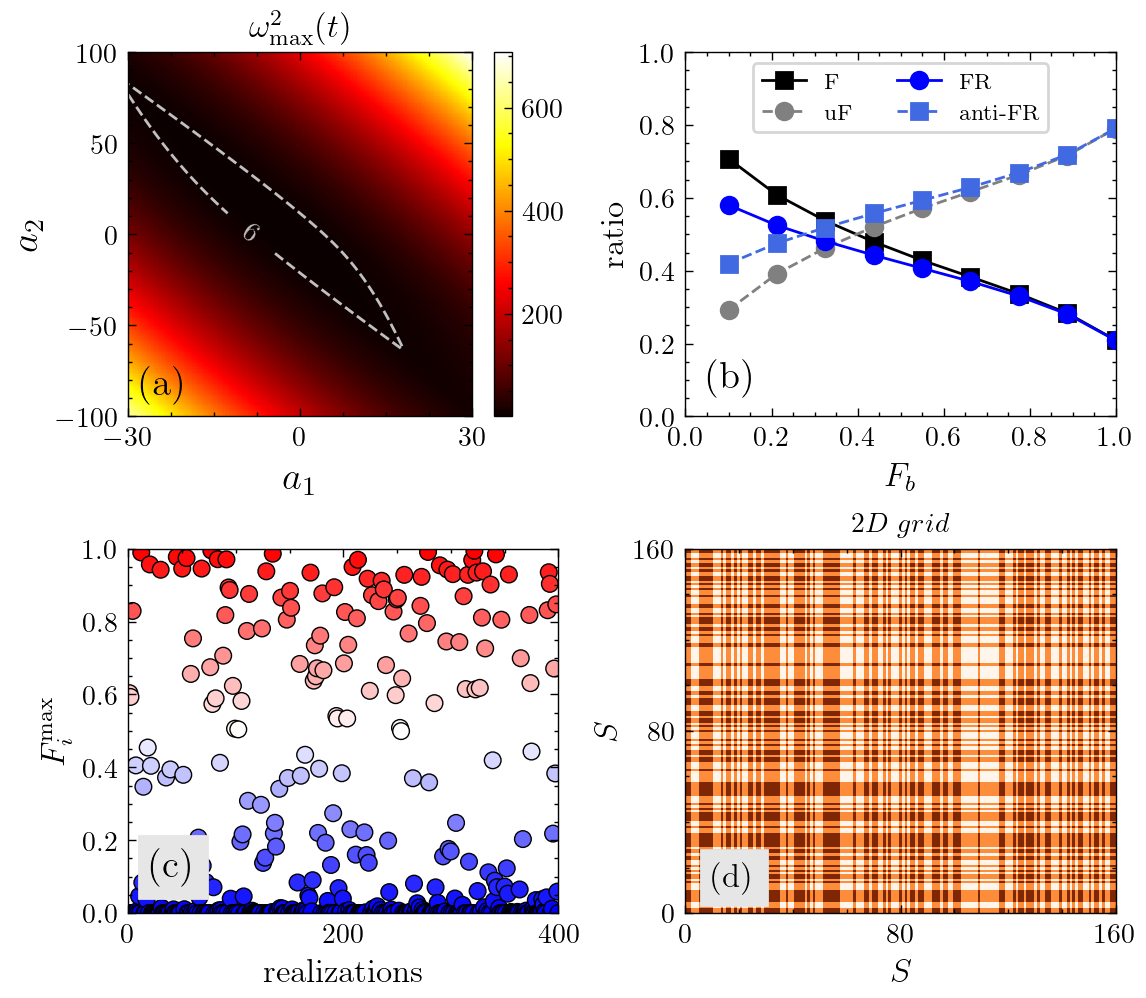}
	\caption{(a) Dependence of the maximum value of $\omega^{2}(t)$ on the coefficient set $\{a_{1},a_{2}\}$, 
		and the white dashed curve presents the contour  of $\omega^{2}(t) \equiv 6$.  (b) The proportion of four classifications, 
		based on two criteria ($F > F_{b}$ and $\omega_{\max}^{2}(t)\leq\Omega^{2}$), 
		in the prepared database as the function of $F_{b}$. The distribution of maximum fidelity $F_{\rm max}$ 
		for  400 exemplified realizations is plotted in (c), and one of realization in 2D grid is illustrated in (d). 
		Here the criteria $F_{b} = 0.9$ and $\Omega = \sqrt{6}$ are used, and other parameters as the 
		same as those in Fig. \ref{figap1}.}
	\label{figap4}
\end{figure}

Next, we generate $4 \times 10^{4}$ realizations of disorder, and thus calculate 
the maximum fidelity $F_{i}^{\max}$ and the corresponding policy $A_{i}$ of  
$200 \times 200$ coefficient grid in the range of $a_{1}\in [-30,30]$ and $a_{2}\in [-100,100]$. 
Here we set two conditions for the optimal control function. First, it has to provide 
a high fidelity for the quantities of interest, i.e.  $F_{\max}>F_{b}$. 
Second,  the corresponding $\omega^{2}(t)$ should correspond to a moderate 
energy consumption required for the transition, implying 
that $\omega_{\max}^{2}(t)$ has not exceeded a certain value $\Omega^{2}$. 
In practice,  by taking into account the 
experimental constrains, such as the limited laser intensity or the gate field in quantum dots, 
we set the control policy,  $\omega_{\max}^{2}(t)\leq\Omega^{2}$, 
where we take $\Omega=\sqrt{6}$ as the critical value for defining ``feasible" policy.
In Fig. \ref{figap4} (a), the contour curve for $\omega_{\max}^{2}(t)=6$ is presented.
Moreover, the optimal policy $A_{\rm opt}$ is constrained by these two conditions: $F_{\max}>F_{b}$ 
and $\omega_{\max}^{2}(t)\le\Omega^{2}$ (labeled by FH). 
The ratio of the database as a function of $F_{b}$ is also presented in  Fig. \ref{figap4} (b),  from which we find 
that the amount of FH database is decreased when we set larger bound, $F_{b}$, for the fidelity.
Obviously, the disorder effect makes the fidelity worse, though the higher fidelity is desirable in the quantum control in the presence of random environment. 
In order to keep balance between the amount of high-fidelity realizations and the diversity of database, we set the bound $F_{b} = 0.9$ as the criteria for keeping the reasonable database, see  Fig. \ref{figap4} (c).
With the assistance of the prepared database satisfying such criteria, we shall discuss the network and training process  as follows.

Previously, we attempted to find the regression between $S_{i}^{\left[2D\right]}\left[j_{1},j_{2}\right]$
and $A_{\rm opt}$  by using only one CNN.
However,  the results are not reasonable, and a very complex neural network is required to provide 
the expressibility and universality for the variety of disorder realization.
Nevertheless, we create an intuitive scheme to reduce the complexity of database, that is, 
the classification is added  prior to the regression.
The database is divided into two categories by the pretraining process: the realization satisfying 
feasible high-fidelity (labeled FH) criteria or not (labeled anti-FH).
As a consequence,  the database for regression is firstly filtered by the classification (CNN1) 
process based on two aforementioned criteria, and secondly train the network (CNN2) based on previously identified FH database.

%

Now, we train the CNN1 with the input $X=S_{i}^{[2D]}[j_{1},j_{2}]$ 
and output $Y = \{0,1\}$ by selecting the loss 
function ${\tt CrossEntropyLoss()}$ respect to Eq. (\ref{cost_class}).
The  identified FH database from CNN1 is the input data of CNN2, and the output is optimal 
policy ${A_{\rm opt}}$ with the loss function  ${\tt MSELoss()}$ in Eq. (\ref{cost_regress}).
Two architectures of CNNs are presented in Fig. \ref{figap3},  where there are 7 layers in a regular sequential 
network CNN1 and 34-layer  ResNet34 \cite{S1ResNet} for CNN2.
Meanwhile, we use the optimizer ${\tt Adam()}$ \cite{S1adam} to optimize the parameters 
based on the gradient descent algorithm.
Moreover, we define the ${\tt Accuracy}= N_{r}/N$ (with $N_{r}$ being the 
number of the right predictions out of total $ {N}$) for CNN1, which is the correct prediction 
number over the total amount of database.
Meanwhile, for quantifying the result of regression, we also define the fidelity 
deviation $\Delta F= |F^{\max}_{i} - F^{\prime}_{i}|$, where $F^{\max}_{i}$ is the 
actual maximum fidelity and $F^{\prime}_{i}$ is the numerical result from the policy predicted by the network
(as in Sec. \ref{Mlp} of the main text).

To this end, we formulate the training algorithm as 

\begin{algorithm}[H]
	\SetAlgoNoLine  
	\caption{Training CNN for classification/regression}
	\KwIn{The database $\{x,y\}$}
	\KwOut{ Trained CNN}
	
	initialization; \\
	optimizer=Adam(learning rate = 0.0001);\\
	loss = CrossEntropyLoss()(or~MSELoss());\\
	\While{epoch}{
		~~~~\For{batch in range(epoch size)}{
			~~~~~~~~~net.train(),\\
			~~~~~~~~~predications = net($x_{i}[batch]$),\\
			~~~~~~~~~training loss = loss(predications, y[$batch$]),\\
			~~~~~~~~~optimizer(net),\\
			~~~~~~~~~total loss  += training loss,\\
			Average Loss = total loss/epoch size,\\
			$epoch+=1$,
		}
	}\label{algorithm1}
\end{algorithm}

\subsection{machine learning outcome}
\label{AppendixB3}

In this section, we present a detailed training process and further discuss the results.
To proceed with the training and testing, we choose the parameters, such as $\omega_{0} = 1$, $\omega_{f} = 0.1$, and $t_{f} = 1$.
The coefficients in the control function of $\omega(t)$ are 
in the range of $a_{1}\in[-30,30] $ and $a_{2}\in[-100,100] $, and the classification criteria  
are  $F_{b} = 0.9$ and $\Omega=\sqrt{6}$.
The whole database $X=\{S_{i},F_{i}^{\max}, A_{i}\}$ for $4 \times 10^4$ realizations in the  $200 \times 200$ coefficient 
grid are established by a 50-core computer for more than 10 hours.
The input data for CNN1 is a 2D random grid  $x_{i} = \{S_{i}[2D]\}$ and the output is $y_{i} = \{0,1\}$, 
to classify the optimal policy is FH $(y=1)$ or anti-FH $(y =0)$. Remarkably, CNN1 manages 
to select $5886$ realizations out of $4 \times 10^{4}$, when the criteria, $F>F_{b} = 0.9$ and 
$\omega_{\max}^{2}(t)\leq \Omega^{2}=6$, are stipulated. 
Eventually,  we  convert these classified realizations  into the CNN2 as the input database,
and the corresponding optimal policy $A_{\rm opt}$ is obtained as the output data. 
For both two networks, $80\%$ of input database is the training database and the rest testing part.
One can find other parameters in Fig. \ref{figap3} and more details in the code.

\begin{figure}
	\centering
	\includegraphics[width=\columnwidth]{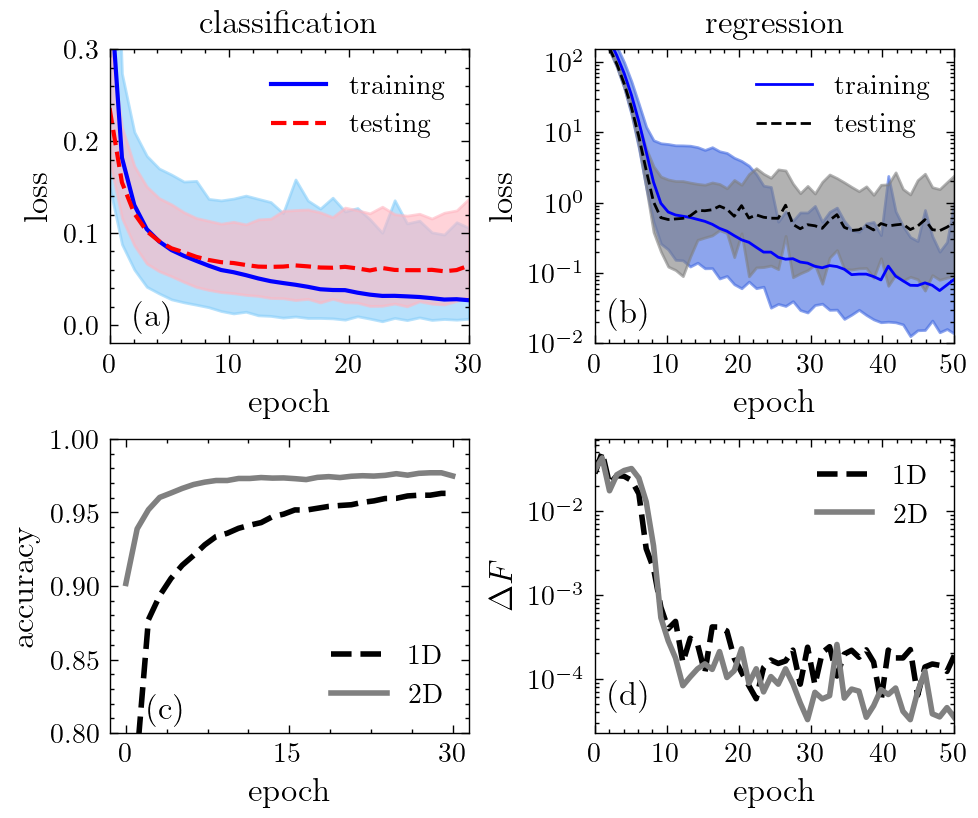}
	\caption{The training and testing loss as the function of epochs in CNN1 (a) and CNN2 (b) for classification and regression. The performances of CNN1 (c) and CNN2 (d) are compared  by using 1D (dashed curve) and 2D (solid curve) input data. Here the shadows are the values of the batches in each epoch. 
	}
	\label{figap5}
\end{figure}

It turns out that the accuracy of CNN1 can reach $97\%$ after $30$ iterations (epoch = $30$), and the fidelity deviation for CNN2 is below than $10^{-4}$ after $50$ iterations (epoch = $50$).
The average loss of training and testing data are presented by the solid and dashed  curves in Fig. \ref{figap5} (a, b), where we see that  the overfitting occurs at $10$ epoch for classification, and  at $20$ epoch for regression.

After that, we shall discuss the generality of our method and the tolerance of model for changing the hyperparameters.
First of all,  we compare the performance of two trained CNNs by using 1D and 2D input data. 
The accuracy and fidelity deviation $\Delta F$ for 1D and 2D input data are presented in Fig. \ref{figap5} (c, d),
where ${\tt Conv1d()}$ and $ {\tt Conv2d()}$  are exploited for 1D
and  2D cases, and the rest parameters are same.
It is evident that the model using 2D input data outperforms the 1D model in terms of accuracy and fidelity deviation.

\begin{figure}
	\centering
	\includegraphics[width=\columnwidth]{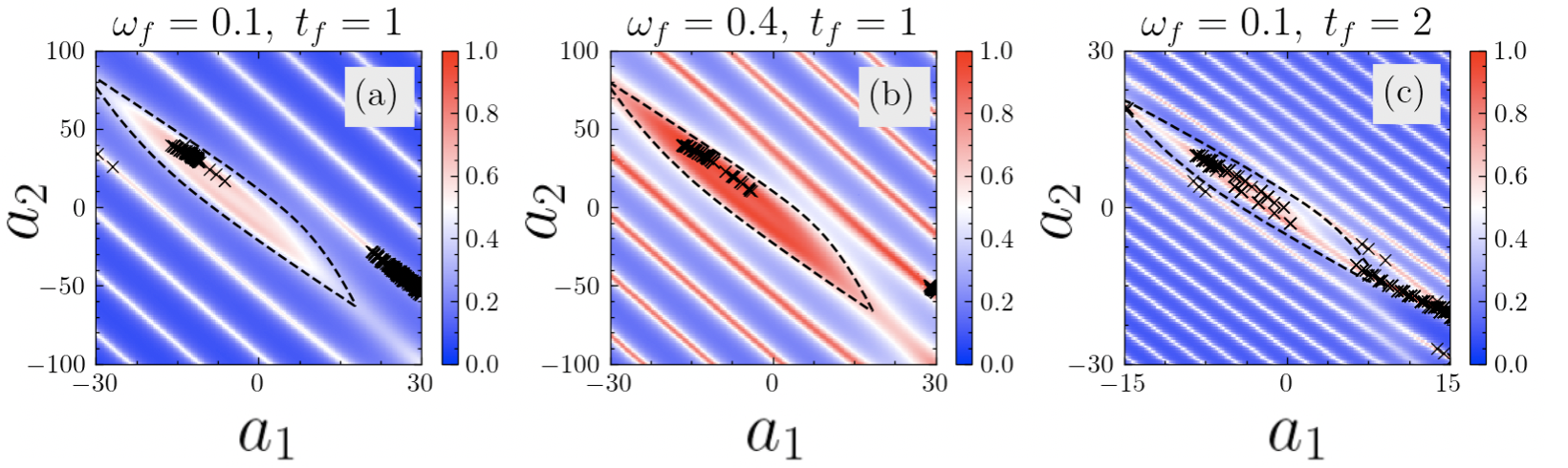}
	\caption{The fidelity as  the function of the coefficient grid $\{a_{1},a_{2}\}$ for various  $\omega_{f}$ and $t_{f}$,
		where  (a) $\omega_{f} = 0.1, t_{f}=1$,  (b) $\omega_{f} = 0.4, t_{f}=1$, and  (c) $\omega_{f} = 0.1, t_{f}=2$ are considered. 
		The location of the maximum fidelity (black cross) is specified for $3.2\times10^4$ realizations of disorder in 
		each plot. The restriction imposed by $\Omega = \sqrt{6}$ is illustrated by black dashed curve in (a, b, c).
		The other parameters are the same as those in Fig. \ref{figap1}. }
	\label{figap6}
\end{figure}

Second, we elaborate the generality of our training model by checking the performance with various values of $\omega_{f}$ and $t_{f}$.
To this end, we prepare the databases of $1.6 \times 10^4$ realizations for $t_{f} =\{1,2,3,4\}$ and $\omega_{f}=\{0.2,0.4,0.6,0.8\}$, 
the criteria and parameters of two CNNs are the same as previous case when $t_{f} = 1, \omega_{f} = 0.1$.
In Fig. \ref{figap6},  we specify the maximum fidelity located in the whole database for the various conditions, 
where  (a) $\omega_{f} = 0.1, t_{f}=1$,  (b) $\omega_{f} = 0.4, t_{f}=1$, and  (c) $\omega_{f} = 0.1, t_{f}=2$ are considered. 
By comparison, the larger $\omega_{f}$ results in the higher fidelity, since the random realizations are much 
easier to recognize, when the final trap frequency is
increased. This is due to the fact that   	
the  influence of random potentials on the fidelity can be negligible, when the final trap potential 
is strong enough such that the localized state has to be located near the origin. 
Consequently,  the  lower loss of CNN1 is achieved since the most of disorder realizations 
are labeled as FH, on the contrary, more inputs causes the performance of CNN2 to degrade. 
In addition, according to the time-energy trade-off, the increase of total time $t_{f}$ makes 
the designed trap frequency easier to satisfy the predetermined criteria ($F>F_{b} =0.9$ and $\omega_{\max}^{2}(t)\leq\Omega^{2}=6$), 
yielding the larger area in Fig. \ref{figap2} (c). In this case,  the database is difficult 
to recognize, see Fig. \ref{figap6}, since more random realizations corresponding 
to the feasible $A_{\rm opt}$ increase the statistical uncertainty and degrade the performance of trained CNNs. 
Therefore, the loss of CNN1 becomes larger when the total time $t_{f}$,
but the loss of CNN2 decreases conversely.
All these results are consistent with 
those of accuracy and fidelity {deviation} in Fig. 5 of main text. Clearly, the quantity and quality 
of the database determine the performance of CNNs, depending  on
the physical constraints or conditions, or the total time, the amplitude of disorder, and trapping potential. 
More important,  we conclude that the interplay of the trapping potential and disorder is of {critical} significance for controlling the dynamics {in terms of the fidelity and the required energy.}

\begin{figure}
	\centering
	\includegraphics[width=\columnwidth]{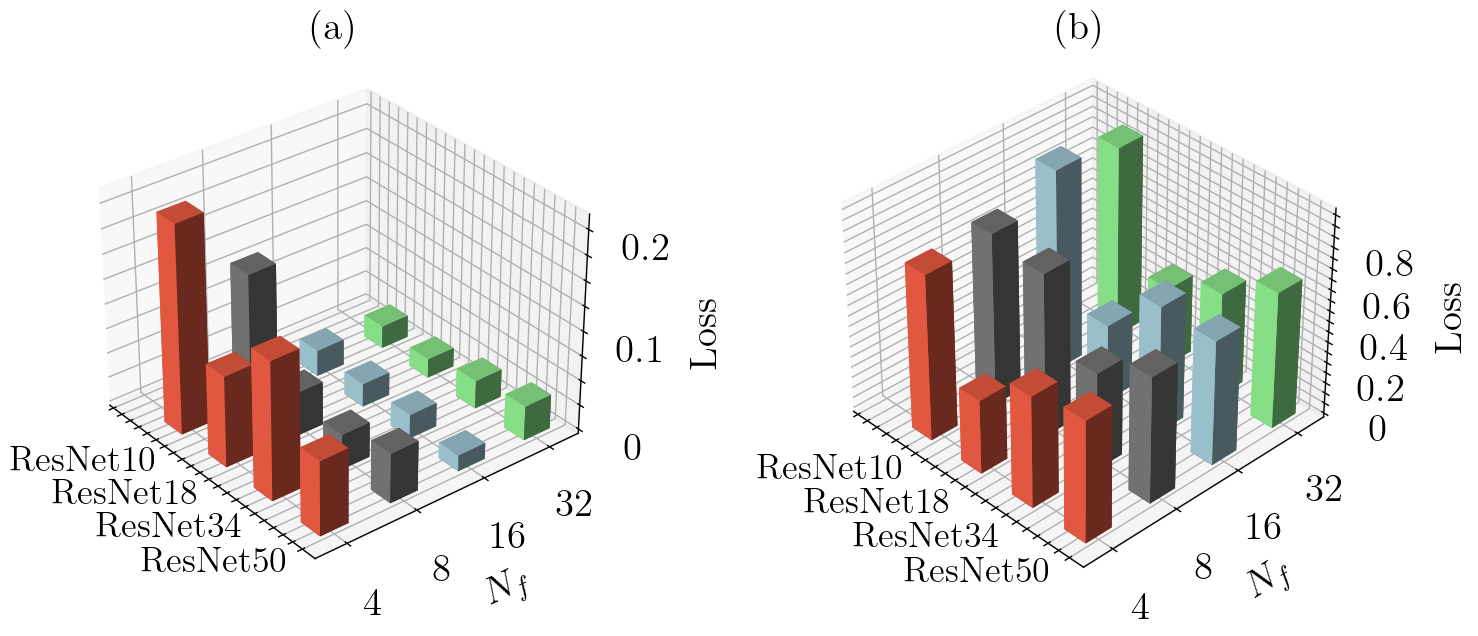}
	\caption{The average loss of training (a) and testing (b) data for different layer number and $N_{f}$ in CNN2. 
		The average loss is the average one of last $10$ epochs among $50$ epochs in the training process. 
		The other parameters are the same as those in Fig. \ref{figap3}.
		Noting we here use another database with the same size of $4\times 10^4$ for clarifying the effect of hyperparameters.
	}
	\label{figap7}
\end{figure}

Finally, we check the performance of the deep CNNs in terms of the hyperparameter, 
such as  the number of hidden layers, the size and number of filters, etc. 
In our model, 
the depth of the CNN2 is much larger than that of CNN1, which suggests that the CNN2 is more sensitive to the hyperparameters.
For simplicity, we concentrate on two hyperparameters, 	the filter number and hidden layers, in the CNN2.
In {this network,} the first layer's outchannel number is $N_{f}$ (see Fig. \ref{figap3}), which determines the total number of filters.
With different $N_{f} = \left[4,8,16,32\right]$, we compare the average loss of testing data for 
10-layer ResNet10, 18-layer ResNet18, 34-layer ResNet34, and 50-layer ResNet50.  
The {clear dependence on} these hyperparameters is presented in Fig. \ref{figap7} (a, b), in which
the corresponding average training and testing losses of the last ten epochs are calculated by 
using  same parameters, respectively. 
Obviously,  the expressibility of network {depends on} the number of parameters.  The average 
training loss {decreases} when the number of layers or filters {increases.}
However, we {emphasize} that the over-fitting of the network appears when the network 
complexity (the number of nodes and alternative paths) 
{increases,} see Fig. \ref{figap7} (b). 
Here we note that all calculations are implemented  by using the online computation resource from the 
Google's cloud service called `$Colab$', which contains GPUs acceleration. For $30$ epochs, 
it takes {about 300 seconds for training the CNN1, but more than $10^{3}$ seconds for the CNN2 while 
	calculation of the fidelity deviation $\Delta F$ takes several hours. The suggested algorithm can be realized at a regular computer 
	without GPU's acceleration albeit with a much longer computation time.}

\section{discussions}
\label{AppendixC}

\subsection{gradient-descent optimization}

\label{AppendixC1}

Here we discuss the generality of the ansatz used here in our proposed method. 
One might be interested to try other ansatz and even optimal (or near-optimal) approach, combined with ML. 
Regarding the latter,  a powerful numerical tool, for 
example, the gradient-descent (GD) algorithm can be applied directly, not as a working tool of the ML algorithms.
To clarify the advantages and disadvantages of this approach, let us study the possible trade-off 
on the improvement of fidelity in the problem of interest and the ability of training CNNs. 
Thus, we compare the optimal solutions produced by GD with the polynomial ansatz-based results.

\begin{figure}
	\centering
	\includegraphics[width=\columnwidth]{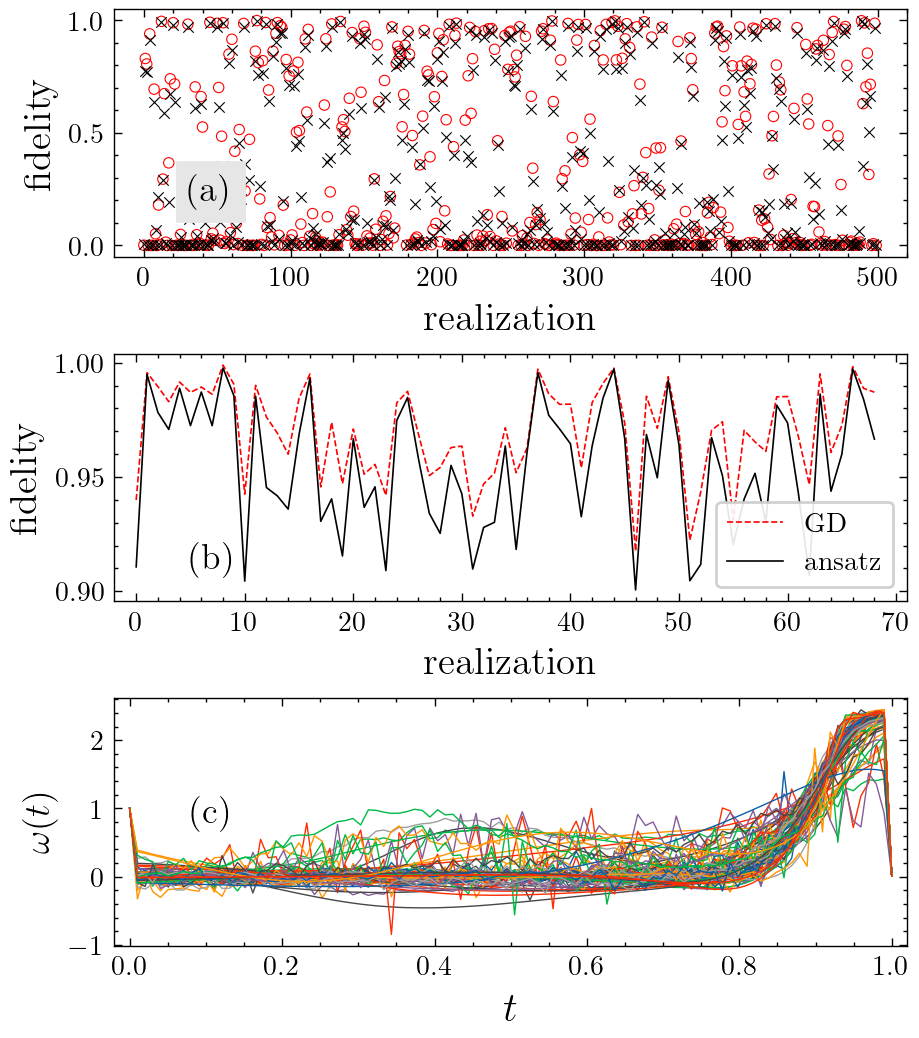}
	\caption{The fidelity distribution for 500 realizations produced by GD-(red circle) and ansatz (black cross)-based scheme in (a), 
		and 69 high-fidelity realizations satisfying high-fidelity $(F>0.9)$ by both methods are illustrated in (b). 
		We present corresponding GD-based control functions in (c).  Parameters: $N_{t} = 100$,
		others that two methods share are the same as those in (\ref{figap1}). Note a similarity between panel (c) and Fig. \ref{fig2}(b), with 
		both figures demonstrating an increase in $\omega(t)$ close to the end of the potential expansion.}
	\label{figap8}
\end{figure}

A parametric optimization problem is the minimization of a given cost-function by gradient descent.
The optimal solution $M^{\rm opt}$ can be produced by minimizing cost-value  $c =J(M)$, which can be expressed as:
\beq
M^{\rm opt} = \min_{c}J(M).
\eeq
In  our scenario, the control function is the trap frequency,  $f(t) =\omega(t) $, with $N_{t}$-intervals 
discrete time $t\in [0, t_{f}]$ (keeping the same $t_{f} = 1$ as that in main text).
Accordingly, the control tuple $f(t) =\{f(0),f(dt),...,f(t_{f})\}$ is constrained by $|f(t)|\le\Omega=\sqrt{6}$ 
and satisfies boundary conditions, e.g. $f(0)=1$ and $f(t_{f}) = 0.1$. 
Then, we shall optimize the $N_{t}$-size tuple $f(t)$ for approaching the highest fidelity by 
minimizing the infidelity $1-F$, in the context of  parametric constrained minimization problem.
In this regard, we perform the optimization process by algorithm $\tt{SLSQP}$ \cite{SLSQP} based on the $\tt{scipy}$ platform.
For one typical realization of random potential, the GD takes several minutes for searching the optimal control function 
which satisfies the convergent condition ($|dJ/dM|<10^{-7}$) of  the cost function while 
our two-step supervised learning method produces the near-optimal solution in several seconds.
Next, we are concerned about the efficiency of  training two CNNs by using GD-produced databases. 

\begin{figure}
	\centering
	\includegraphics[width=\columnwidth]{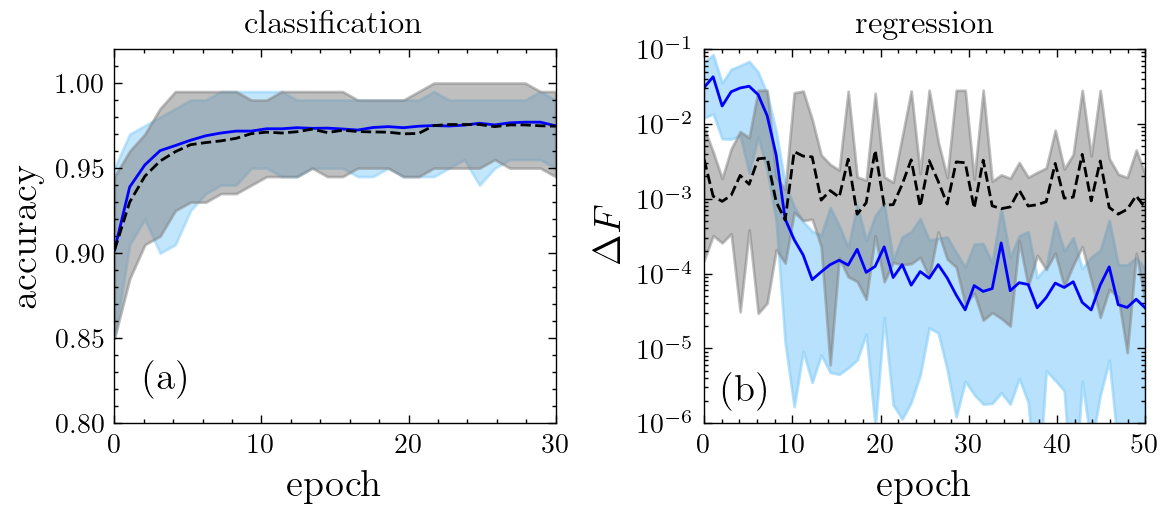}
	\caption{ Databases generated for training two CNNs by two techniques: GD (black dashed) and ansatz-based (blue solid).
		Left: the accuracy of classification for testing data versus training epoch. 
		Right: $\Delta F$ of testing data versus training epoch for regression. In both sub-figures, 
		the corresponding shadowed area contains the result of training batches in each epoch, 
		and curves are the average values. The shared parameters are the same as those in the main text. }
	\label{figap9}
\end{figure}

To this end, we 
calculate the GD-based control function for the same $4\times\,10^4$-realization database used in the main text.
It is expected that the GD method with $N_{t} = 100$ improves the fidelity. Thus, it increases the number of FH realizations, thus providing 6801 of them
against 5886 for ansatz-based method.
In Fig. \ref{figap8} we present the fidelity distribution of 500 realizations in (a) for two methods: GD (red circle) and ansatz-based (black cross).
More distinctly, we compare 69 realizations among 500, which admit the high fidelity for both methods in (b) of Fig. \ref{figap8},
with the corresponding 69 optimal solutions produced by GD illustrated in (c).
One can see that, although the GD-based method slightly increases the fidelity compared to the ansatz-based one, it does not 
change the fidelity distribution strongly. This result can be understood by the physics argument that the fidelity of control policy depends mainly on the localization induced by random potential 
rather than on  the control strategy. 
Figure \ref{figap9} further demonstrates the performance of CNN1 (classification) and CNN2 (regression) trained by the two databases generated from a simple ansatz (blue solid) and GD (black dashed).
In addition,  one can see the disadvantages of GD-based optimal control as the database for training CNN2.
The GD method indeed boosts the fidelity of control policy on the cost of loosing the generality in CNN2.
It is due to the fact that the performance of GD-based CNN2 is worse: the database dimension $N_{t} = 100$ is much 
larger than that for the ansatz-based database (which is 2), eventually decreasing the reliability of the regression process.
Thus,  the  balance between the fidelity improvement and the ability to train the CNN should be kept as our method.

\begin{figure}
	\centering
	\includegraphics[width=\columnwidth]{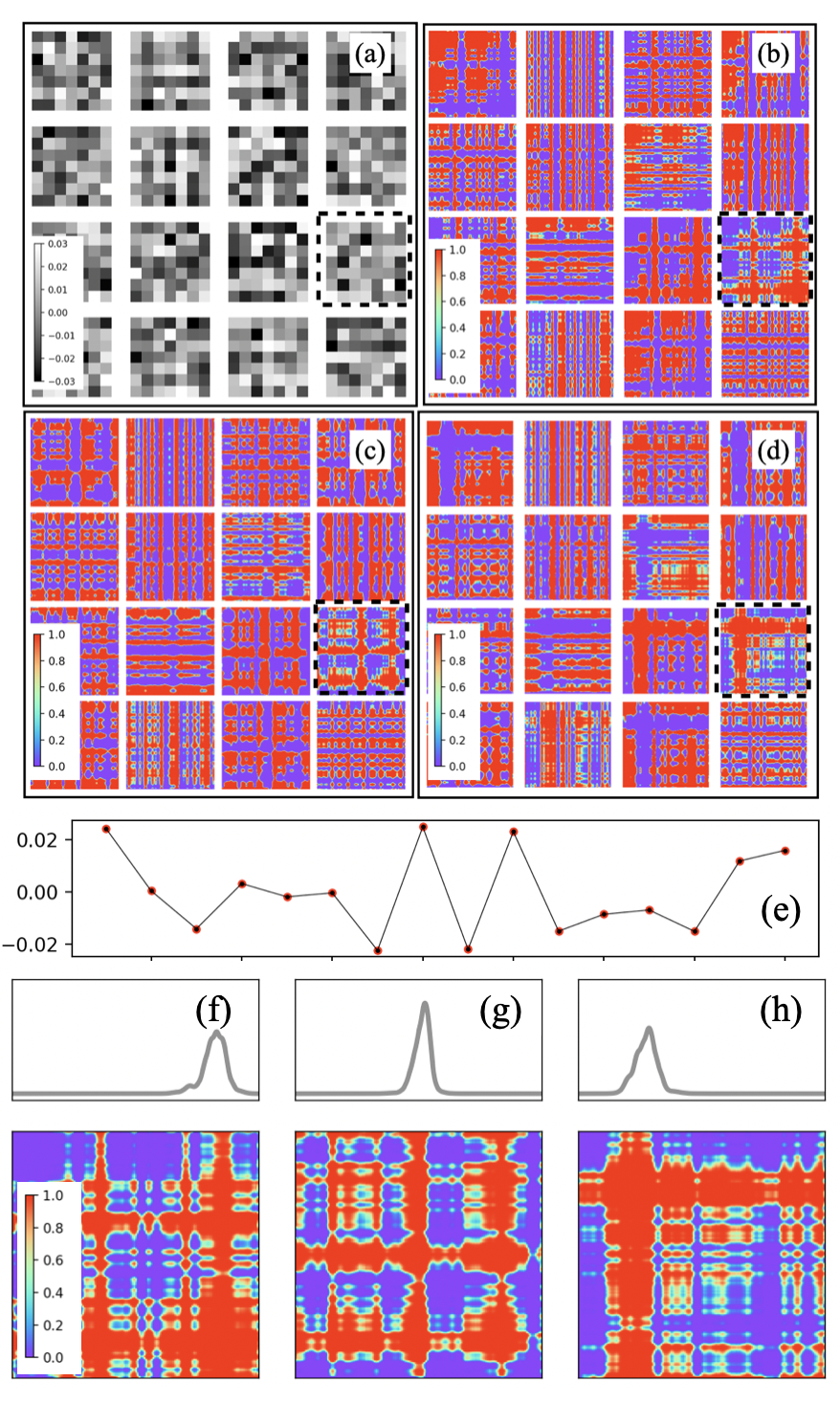}%
	\caption{16 parametric filters of 4-th layer for CNN1 in (a) and related bias in (e) . (b-d): the feature-map for three different realizations, and corresponding densities of wave-packets and the selected feature-map (labeled by black dashed squares ) in (f-h), respectively. }
	\label{figap10}
\end{figure}

\begin{figure}
	\centering
	\includegraphics[width=\columnwidth]{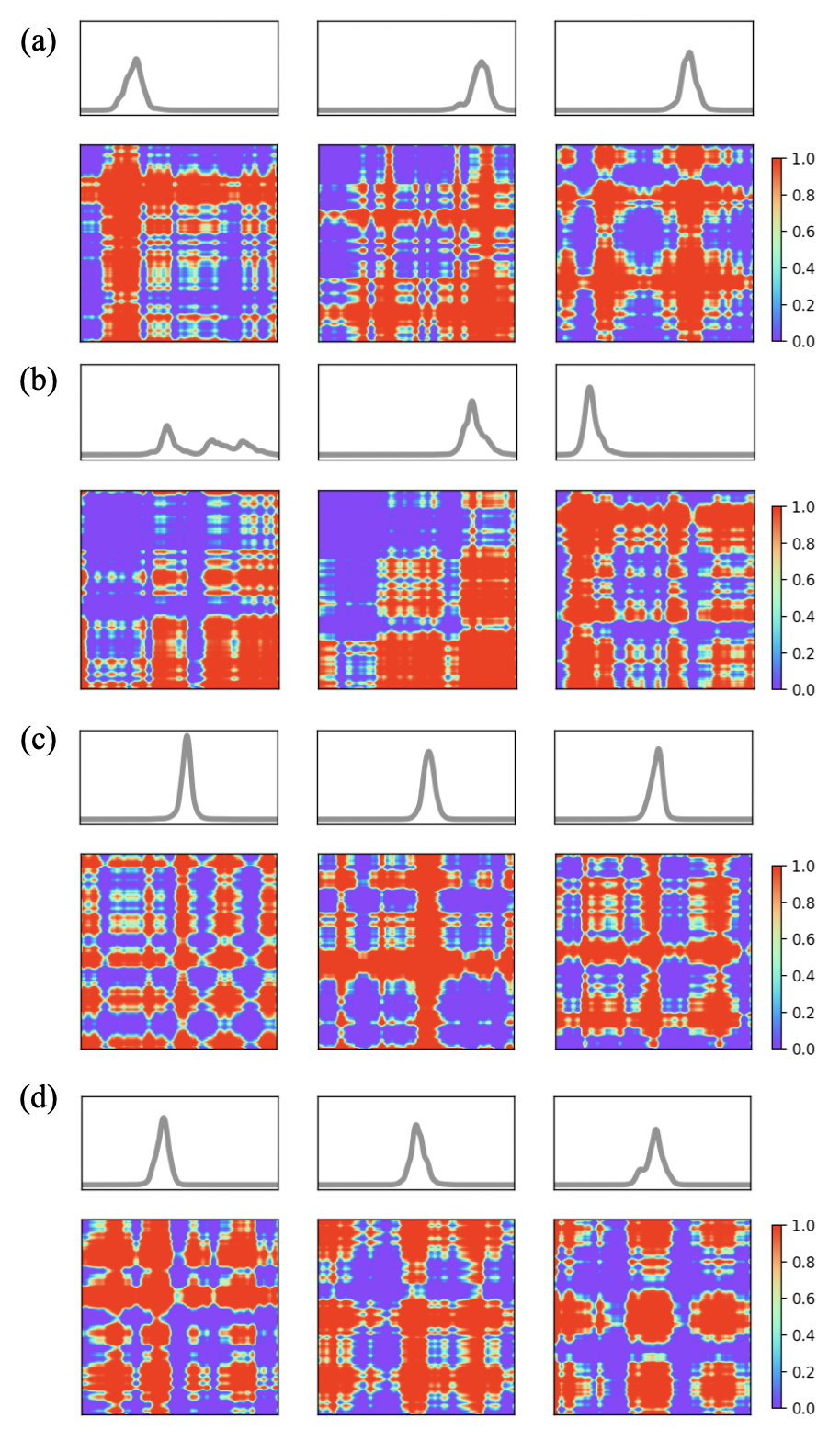}
	\caption{The wave-packet density and related feature-map selected as in Fig. \ref{figap10} for 12 realizations. Panel (a-b) for low-fidelity $(F<0.9)$ and (c-d) for high-fidelity $(F>0.9)$ cases.  }
	\label{figap11}
\end{figure}

\subsection{interpretability of CNNs}
\label{AppendixC2}

{It is difficult to explain the results obtained from ML in an intuitive way, despite of many successful applications in quantum physics  \cite{krizhevsky2012imagenet,CarrasquillaNature2017}. In order to understand the machine-making decision in solving the optimal control problem,
	we discuss the interpretability (or explainability) of a ML task.
	The interpretability in ML is defined, for example, by Miller \cite{Miller}: 
	\textit{'Interpretability is the degree to which a human 
		can understand the cause of a decision'} or, similarly, by Kim \cite{kim} as: 
	\textit{'Interpretability is the degree to which a human can constantly predict the model's result'.} 
	The interpretability of a training model brings criteria such as comprehensibility, reliability, and 
	fairness of facts upon the process of ML.  In a recent work \cite{molnar2020interpretable}, Molnar offers a comprehensive review on the concept, 
	principles and importance of explainable models in the field of ML.
	Among them, we offer here the evidence of interpretability 
	by visualizing the feature- map of CNN1 for understanding and explaining the ML outcomes \cite{molnar2020interpretable}.}

{First, we recall the element of output from the convolution operation ${\tt Conv2d()}$  : 
	\beq
	x^{\ell}_{i,j} = \sum_{m,n}^{k_{1},k_{2}} K_{m,n}^{k}x_{i+m,j+n}^{\ell-1}+b^{\ell},
	\eeq
	where the $\ell$-th feature-map $x^{\ell}_{i,j} $ is the sum of  product of filters $K_{m,n}^{k}$, and 
	corresponding filter-size $(l-1)$-th feature-map $x_{i+m,j+n}^{\ell-1}$ with bias $b^{\ell}$.
	According to the structure of CNN1 designed in Fig. \ref{figap3}, we have sixteen $ 7\times 7$ 
	weight matrices (filters) in each convolution layer.
	In Fig. \ref{figap10},
	we present 16 parametric filters of  last layer in (a) and corresponding bias in (e), 
	and produce 16 feature-maps for three selected realizations in (b-d) after ${\tt Sigmoid}$ function.
	For illustration, 
	we extract the most representative feature-maps (labeled by black dashed squares) out of 16 in (e-h) of Fig. \ref{figap10}, and compare them with the corresponding density of the final wavepacket 
	with the trap frequency $\omega(t_{f})=\omega_{f}$.
	By performing the 4-layer convolution product operation, an original input 
	2D random grid (see Fig. \ref{figap4}) is transformed into a particular feature map,
	which can be interpreted by the localization of the target state density.
	More specifically, the feature-map is strongly correlated with the localization of 
	density for low-fidelity realization, such as (e) and (g) in Fig. \ref{figap10}.
	For the high-fidelity case, the feature-map is much more uniformly distributed 
	compared to the low-fidelity counterparts.
	In Fig. \ref{figap11}, we further compare the final state probability density and feature-map for 12 realizations 
	including low-fidelity (a-b) and high-fidelity(b-d) realizations.
	To this end, one can't precisely identify the random sequence just by watching the feature-map,
	in particular, for realizations with $F_{b}$ close to neither 1 nor 0.  
	However, for realizations with fidelity $F\ll\,1$ or $F\to 1$ can be easily identified and 
	explained, according to the typical feature-map in (e-g) of Fig. \ref{figap10}.
	It should be emphasized that  our results can be interpreted based on the
	comparison of the feature-map and the wavepacket density.
	In a word,  the accurate ML outcome captures the hints from the feature-maps, which are related to the nature of the localization physics in random potentials. 
}

\bibliography{ref}

\end{document}